\documentclass[amsmath,amssymb,superscriptaddress,showpacs,twocolumn]{revtex4-2}

\usepackage{graphicx}
\usepackage{dcolumn}
\usepackage{bm}
\usepackage[colorlinks=true, citecolor=blue, urlcolor = blue, linkcolor= red, bookmarks=true]{hyperref}

\begin{document}
\title{Rotating Black Holes in Horndeski Gravity: Thermodynamic and Gravitational Lensing}
\author{Rahul Kumar Walia}\email{rahul.phy3@gmail.com}
\affiliation{Astrophysics Research Centre, School of Mathematics, Statistics and Computer Science, University of
	KwaZulu-Natal, Private Bag 54001, Durban 4000, South Africa}
	\author{Sunil D. Maharaj}
\email{maharaj@ukzn.ac.za}
\affiliation{Astrophysics Research Centre, School of Mathematics, Statistics and Computer Science, University of
	KwaZulu-Natal, Private Bag 54001, Durban 4000, South Africa}
\author{Sushant G. Ghosh} \email{sghosh2@jmi.ac.in, sgghosh@gmail.com}
\affiliation{Astrophysics Research Centre, School of Mathematics, Statistics and Computer Science, University of KwaZulu-Natal, Private Bag 54001, Durban 4000, South Africa}
\affiliation{Centre for Theoretical Physics, Jamia Millia Islamia, New Delhi 110025, India}

\date{\today}

\begin{abstract}
	The lack of rotating black holes, typically found in nature, hinders testing modified gravity from astrophysical observations. We present the axially symmetric counterpart of an existing spherical hairy black hole in Horndeski gravity having additional deviation parameter $Q$, which encompasses the Kerr black hole  as a particular case ($Q=0$). We investigate the effect of Horndeski parameter $Q$ on the rotating black holes geometry and analytically deduce the gravitational deflection angle of light in the weak-field limit. For the S2 source star, the deflection angle for the Sgr A* model of rotating Horndeski gravity black hole for both prograde and retrograde photons is larger than the Kerr black hole values. We show how parameter $Q$ could be constrained by the astrophysical implications of the lensing by this object.  The thermodynamic quantities, Komar mass, and Komar angular momentum gets corrected by the parameter $Q$, but the Smarr relation $M_{\text{eff}}=2ST+2\Omega J_{\text{eff}}$ still holds at the event horizon.  
\end{abstract}


\maketitle
\section{Introduction}
The first exact solution to Einstein's general relativity (GR) was found by Schwarzschild \cite{Schwarzschild:1916uq}, describing a static, spherically symmetric black hole. However, the rotating Kerr \cite{Kerr:1963ud}, and Kerr-Newman metrics \cite{Newman:1965tw}, which are astrophysically more relevant, represent rotating black holes that arise as the final state of gravitational collapse. As predicted by GR, black holes now have direct observational evidence of their existence after the Event Horizon Telescope (EHT) collaboration released the first image of the supermassive black hole M87* \cite{Akiyama:2019cqa, Akiyama:2019bqs}. Although the  EHT observation of M87* is consistent with the Kerr black hole's image as predicted by GR, the observation did not say anything about most modified gravity theories or alternatives to the Kerr black hole. Lately, with the M87* shadow angular size, constraints are placed on the second post-Newtonian metric coefficients, which were inaccessible in the earlier weak field tests at the solar scale \cite{EventHorizonTelescope:2020qrl}, and on the physical charges of a large variety of modified gravity black holes  \cite{EventHorizonTelescope:2021dqv,Bambi1,Medeiros:2019cde}. 

Also, the recent cosmological observations are consistent with standard models built on GR but imply a mysterious late-time acceleration phase. Although the cosmological constant is the most straightforward candidate to drive the late-time acceleration, it poses an insurmountable problem to quantum field theory since it is difficult to accommodate its observed value with vacuum energy calculations \cite{weinberg}. Recent studies suggested that to simultaneously relieve both the cosmological tension $H_0$ and $\sigma_8$ tension one require the equation of state parameter for dark energy $\omega$ to cross the phantom side \cite{Lee:2022cyh,Heisenberg:2022gqk,Heisenberg:2022lob}. Nevertheless, the GR modification on cosmological scales can explain the late-time acceleration. Several modified gravity models can be reformulated in terms of scalar-tensor theories of gravitation; i.e., they are mathematically equivalent to a gravitational theory whose degrees of freedom are the metric $g_{\mu \nu}$ and one or more scalar fields $\phi$. Scalar-tensor theories are probably the simplest, most consistent, and non-trivial modification of GR \cite{Damour}. One of the most famous theories of the scalar-tensor family is Horndeski gravity \cite{Horndeski:1974wa}, which is probably the most general four-dimensional scalar-tensor theory with equations of motion containing second-order derivatives of the dynamical fields. Horndeski theories have been playing a significant role \cite{kase} in describing the accelerated expansion (see \cite{koba} for a review). Horndeski required a theory with second-order field equations \cite{Horndeski:1974wa}. His proposed approach results in a theory that avoids Ostrogradski instabilities \cite{Ostrogradsky}, and it is a priori eligible to have a ghost-free vacuum. Further physical relevance of such models could be tested, particularly in the strong-field regime, namely around black holes. The Horndeski theory of gravity in four dimensions has similar features to GR, namely diffeomorphism invariance and second-order field equations. Interestingly, Horndeski gravity can be obtained from higher dimensional Lovelock theory with the help of Kaluza-Klein procedure \cite{Lu:2020iav,VanAcoleyen:2011mj,Charmousis:2014mia}, such that its space of solutions is endowed with hairy black holes \cite{Rinaldi:2012vy, Babichev:2014fka,Anabalon:2013oea,Cisterna:2014nua,Bravo-Gaete:2014haa}. Black hole solutions with new global charges, which are not associated with the Gauss law \cite{Herdeiro:2015waa}, are called hairy black holes, e.g., black holes with scalar hair \cite{Herdeiro:2014goa,Gao:2021luq}, or proca hair \cite{Herdeiro:2016tmi}. See Ref. \cite{Herdeiro:2015waa} for a recent review on black holes with hair due to global charge. 

Hui and Nicolis \cite{Hui:2012qt} demonstrated that the no-hair theorems do not apply when a Galileon field is coupled to gravity because of Galileon's field peculiar derivative interactions. Nonetheless, they showed that static spherically symmetric black holes could not sustain nontrivial Galileon profiles. The authors \cite{Babichev:2017guv} further examined the no-hair theorem in Ref.~\cite{Hui:2012qt} considering Horndeski theories and beyond. They demonstrated that shift-symmetric Horndeski and beyond Horndeski theories allow for static and asymptotically flat black holes with a static scalar field such that the Noether current associated with the shift symmetry vanishes, while the scalar field cannot be trivial; it leads to hairy black hole solutions in quartic Horndeski gravity \cite{Babichev:2017guv}. They considered the Lagrangian of Horndeski theory which can be written as a generalized Galileon Lagrangian, and the corresponding action \cite{Babichev:2017guv}
reads
\begin{align}
S=&\int \sqrt{-g}\Big\{Q_2(\chi)+Q_3(\chi)\Box\phi+ Q_4(\chi) R
+ Q_{4,\chi}\Big[(\Box \phi)^2\nonumber  \\ 
&-(\nabla^\mu\nabla^\nu \phi)(\nabla_\mu\nabla_\nu \phi)\Big] +Q_5(\chi)G_{\mu\nu} \nabla^\mu\nabla^\nu \phi \nonumber  \\ 
& -\frac{1}{6}Q_{5,\chi}\Big[(\Box \phi)^3
-3(\Box \phi) (\nabla^\mu\nabla^\nu \phi)(\nabla_\mu\nabla_\nu \phi)\nonumber  \\ 
&
+2(\nabla_\mu\nabla_\nu\phi)(\nabla^\nu\nabla^\gamma \phi)(\nabla_\gamma\nabla^\mu \phi)\Big]
\Big\} d^4x.	\label{action}
\end{align}
where $Q_2$, $Q_3$, $Q_4$, $Q_5$ are arbitrary functions of the scalar field $\phi$ and $\chi=- \partial^\mu \phi \partial_\mu \phi/2$ is the canonical kinetic term. Additionally, in our notation, $f_{\chi}$ stands for $\partial f(\chi)/\partial \chi$, $R$ is the Ricci scalar, $G_{\mu\nu}$ is the Einstein tensor, 
and 
\begin{eqnarray}\label{nable}	
&& (\nabla_\mu\nabla_\nu\phi)^2 \equiv \nabla_\mu\nabla_\nu\phi \nabla^\nu\nabla^\mu\phi, \nonumber \\ 
&& (\nabla_\mu\nabla_\nu\phi)^3 \equiv \nabla_\mu\nabla_\nu\phi \nabla^\nu\nabla^\rho\phi \nabla_\rho\nabla^\mu\phi.
\end{eqnarray}
The scalar field admits the Galilean shift symmetry $\partial_{\mu}\phi\to \partial_{\mu}\phi+ b_{\mu} $ in flat spacetime for $Q_2\sim Q_3 \sim \chi$ and $Q_4\sim Q_5 \sim \chi^2$, which resembles the Galilean symmetry~\cite{Nicolis:2008in}. 
The simultaneous detection of gravitational waves GW170817 from a binary neutron star merger and their electromagnetic counterpart gamma-ray burst GRB 170817A restrict the functional form of the Horndeski gravity theory as $Q2(\chi)+Q3(\phi, \chi)\box \phi +Q4(\phi)R$ \cite{Kase:2018aps}. On the other hand, some specific Horndeski gravity models with the quartic-order derivative and quintic-order couplings are also consistent with the GWs observations \cite{Amendola:2018ltt}. Moreover, it is true that GW astronomy severely narrowed down the theoretical space for scalar-tensor theories; similar constraints in the electromagnetic spectrum are largely missing. With this assertion, it is interesting and important to investigate to what extent the constraints placed on the Horndeski gravity theories from the GW observations complement those inferred from the electromagnetic spectrum using gravitational lensing and shadow. The EHT probe length scales (curvature scales) are eight orders of magnitude larger (16 orders of magnitude smaller) than those measured by LIGO. In this paper, we started with the generalized Horndeski gravity theory without any prior constraints from GWs observations.

Henceforth, we shall restrict ourselves to a scalar field $\phi\equiv\phi(r)$, which source the static and spherically symmetric spacetime. Recently, the authors in Refs.~\cite{Bergliaffa:2021diw} considered the particular case of Horndeski scalar field viz., $Q_2\propto  (\phi_{,\mu}\phi^{,\mu})^{3/2}$, $Q_3=0$, $Q_4\propto (\phi_{,\mu}\phi^{,\mu})^{1/2}$ and $Q_5=0$ and obtained an exact hairy black hole solution given by 
\begin{eqnarray}
\label{BHS}
ds^2=-A(r)dt^2+\frac{1}{B(r)}dr^2+r^2(d\theta^2+\sin^2\theta d\varphi^2),
\end{eqnarray}
with
\begin{eqnarray}
\label{eqf}
A(r)=B(r)=1 -\frac{2M}{r} + \frac{Q}{r}\ln\left({\frac{r}{r_0}}\right).
\end{eqnarray}
At large distances, $r>> 2M$, the scalar field $\phi\propto \ln(r)$. Here, $M$ is the constant related to the black hole mass, and $Q$ is the hairy parameter coming from the Horndeski theory. The solution (\ref{BHS}) is a class of hairy black holes which encompasses the Schwarzschild solution when $Q \to 0$, and is in fact asymptotically flat since $\lim_{r\rightarrow\infty}A(r)=B(r)=1$. The strong gravitational lensing of the spherically symmetric black hole in Horndeski gravity is investigated in Ref.~\cite{Kumar:2021cyl}. However, the nonrotating black hole solutions are not favourable for observations, as astrophysical black holes rotate, and their spin is crucial for the astrophysical processes. The lack of rotating black hole models in modified gravity theories substantially hinders testing modified gravity theories from observations. Despite its appealing features and numerous applications, rotating solutions in Horndeski gravity are lacking.
Furthermore, the study of gravitational lensing in scalar fields is relatively unexplored, except for a few nonrotating models \cite{Kumar:2021cyl,Virbhadra:1998dy}. It motivated us to seek an axisymmetric generalization of the metric (\ref{BHS}), namely, a rotating Horndesky hairy black hole metric, and to test it with astrophysical observations through gravitational lensing. The Kerr metric \cite{Kerr:1963ud} is the essential general relativity solution representing an exterior vacuum rotating black hole spacetime resulting from gravitational collapse. However, modified gravity theories admit non-Kerr black hole solutions as well. This paper analyses the causal structure of the rotating black hole in Horndeski gravity, discusses the horizon structure and energy conditions and obtains the associated Komar conserved quantities. Furthermore, using the Gauss-Bonnet theorem, we analytically derive the deflection angle in the weak field limit caused by the rotating black holes. We also estimate the correction in the deflection angle due to the hair $Q$ for the supermassive black hole Sgr A* at the Galactic centre.

The organization of this paper is as follows. In Sec.~\ref{sec2}, we construct the rotating counterpart of the hairy black hole (\ref{BHS}),  namely, the rotating Horndeski black hole metric using the revised Newman-Janis algorithm, and also discuss horizon structures and weak energy conditions. In Sec.~\ref{sec3}, we use the rotating spacetime symmetries to derive the conserved mass and angular momentum of the rotating Horndeski black hole and show that the Komar conserved quantity corresponding to the null Killing vector at the horizon obeys $\mathcal{K}_{\chi}=2S_+T_+$. In Sec.~\ref{sec4}, we discuss the gravitational deflection of light in the stationary spacetime. Further, in this section, considering the rotating Horndeski gravity black hole as the Sgr A* black hole, we numerically compute the deflection angle and estimate the corrections from the Kerr black hole. Finally, we summarize our main findings in Sec.~\ref{sec5}.

\begin{figure}
	\includegraphics[scale=0.9]{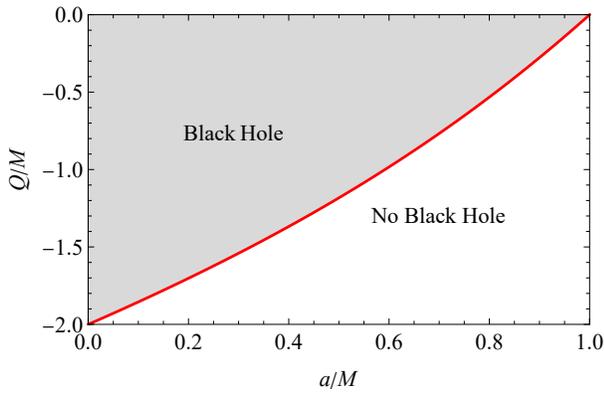}	
	\caption{The parameter plane for $(a,\;Q)$ of the rotating metric, and the thin line separates black holes from configurations without an event horizon.} \label{plot1}	
\end{figure}
\section{Rotating Metric}\label{sec2}
The Newman$-$Janis algorithm (NJA) provided a novel way to obtain rotating spacetimes from a static, spherically symmetric and seed metric, without needing to integrate any of the field equations \cite{Newman:1965tw}. NJA successfully constructs the Kerr (Kerr-Newman) solution from the Schwarzschild (Reissner-Nordstrom) black hole as a solution-generating method \cite{Newman:1965my}. In this paper, starting from a spherical black hole in Horndeski gravity \cite{Bergliaffa:2021diw}, we construct a rotating spacetime using the revised NJA algorithm \cite{Azreg-Ainou:2014pra,Azreg-Ainou:2014aqa}.  The advantage of using revised NJA is that the resulting rotating metric is always expressible in Boyer–Lindquist form, i.e., it is always possible to make a global coordinate transformation from null coordinates to the Boyer$-$Lindquist coordinates.  Furthermore, the revised NJA has been successfully applied to generate rotating black holes in modified gravities \cite{Johannsen:2011dh, Bambi:2013ufa, Ghosh:2014pba, Moffat:2014aja,Kumar:2017qws,Kumar:2020hgm,Kumar:2020owy}. Also, the first-ever rotating black hole solution in the loop quantum gravity is also obtained using the revised NJA \cite{Brahma:2020eos}.  The metric of the rotating hairy black holes spacetime, obtained using revised NJA from spherical black holes (\ref{BHS}), in the Boyer-Lindquist coordinates reads
\begin{align}\label{rotbhtr}
ds^2=& - \left( 1- \frac{2M(r)r}{\Sigma } \right) dt^2 +
\frac{\Sigma}{\Delta }dr^2 + \Sigma d \theta^2 \nonumber  \\ 
&- \frac{4aM(r) r
}{\Sigma } \sin^2 \theta dt \; d\phi + \frac{\mathrm{A}}{\Sigma}\sin^2 \theta d\phi^2,
\end{align}
with
\begin{eqnarray*}
	& & \Sigma=	r^2 + a^2 \cos^2\theta, \;\;\; \Delta=r^2 + a^2 - 2 r M(r) ,\;\; \nonumber \\ 
	& & M(r) = M - \frac{Q}{2 } \ln \left(\frac{r}{r_0}\right), \; \nonumber \\
	& &\mbox{and}\;\; \mathrm{A}= (r^2+a^2)^2 - a^2 \Delta \sin^2 \theta,
\end{eqnarray*} 
in which the spin parameter $a$ is included through revised NJA.
Moreover, we shall demonstrate that the metric (\ref{rotbhtr}), for some values of $Q$ and $a$, can describe rotating black holes up to certain values of the spin parameter $ a $ (cf. Fig. \ref{plot1}) and has a rich structure. The metric~(\ref{rotbhtr}), in the limit $Q\to 0$, goes over to the Kerr black hole \cite{Kerr:1963ud} and also to the spherically symmetric hairy black hole solution (\ref{BHS}) when only $a=0$, and the Schwarzschild solution for $Q=a=0$. It is not difficult to find a range of $a$ and $Q$ for which the solution (\ref{rotbhtr}) is a black hole, as shown in gray region of Fig.~\ref{plot1}. Henceforth, for definiteness, we shall address the solution (\ref{rotbhtr}) as the rotating hairy black holes. 
The rotating hairy black hole metric~(\ref{rotbhtr}), likewise the Kerr black hole, is independent of $\phi$ and $t$ coordinates, and hence admit two Killing vectors, respectively, $\eta_{(t)}^{\mu}=\delta^{\mu}_t$ and $\eta_{(\phi)}^{\mu}=\delta^{\mu}_{\phi}$ with $\delta^{\mu}_{a}$ the Kronecker delta. Thus, by the definition of Killing vectors, four-momentum components associated with translation along $t$ and $\phi$ coordinates are constants of motion for free particle geodesics. The solution (\ref{rotbhtr}) is singular at $ \Sigma = 0 $ and at $\Delta=0$. The surface $\Sigma=0$ is a ring shape curvature singularity of radius $a$ and $\Delta=0$ is a coordinate singularity which determines the horizon -- the horizons of the rotating hairy black hole metric ~(\ref{rotbhtr}) are thus solution of 
\begin{equation}\label{horizon}
g^{\mu\nu}\partial _{\mu}r\partial_{\nu}r=0\;\;\; or\;\; g^{rr}=\Delta=0, 
\end{equation} 
where $\partial _{\mu}r$ is the normal to the said hypersurface. The event horizon is a null surface representing the locus of outgoing future-directed null geodesic rays that never manage to reach arbitrarily large distances from the black hole \cite{Hawking:1971vc,he,Poisson:2009pwt}. 
\begin{figure*}[t]
	\begin{center}	
		\begin{tabular}{c c} 
			\includegraphics[scale=0.85]{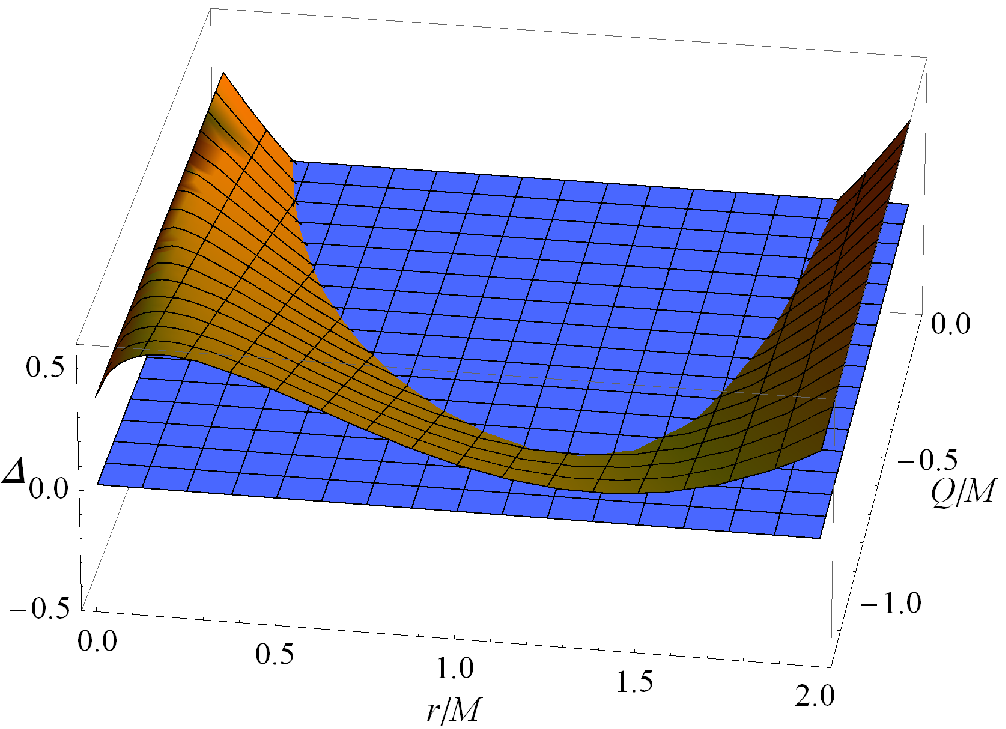}&
			\includegraphics[scale=0.85]{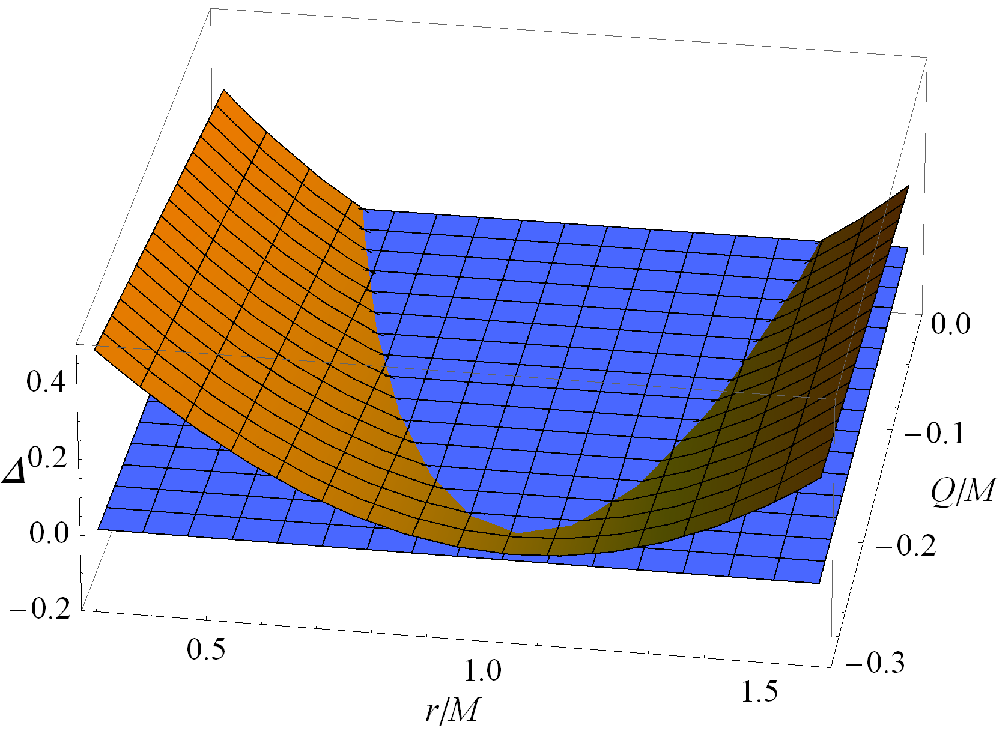}
		\end{tabular}
	\end{center} 
	\caption{Plot showing the variation of $\Delta$ with $r$ for different values of parameters $Q$ for $a=0.6M$ (left) and $ a=0.9M$ (right). The blue surface correspond to $\Delta=0$ and values at the intersection points between two surfaces are the radii of horizons.}\label{plot2}	
\end{figure*}
\begin{figure*}[htb]
	\begin{tabular}{c c}
		\includegraphics[scale=0.85]{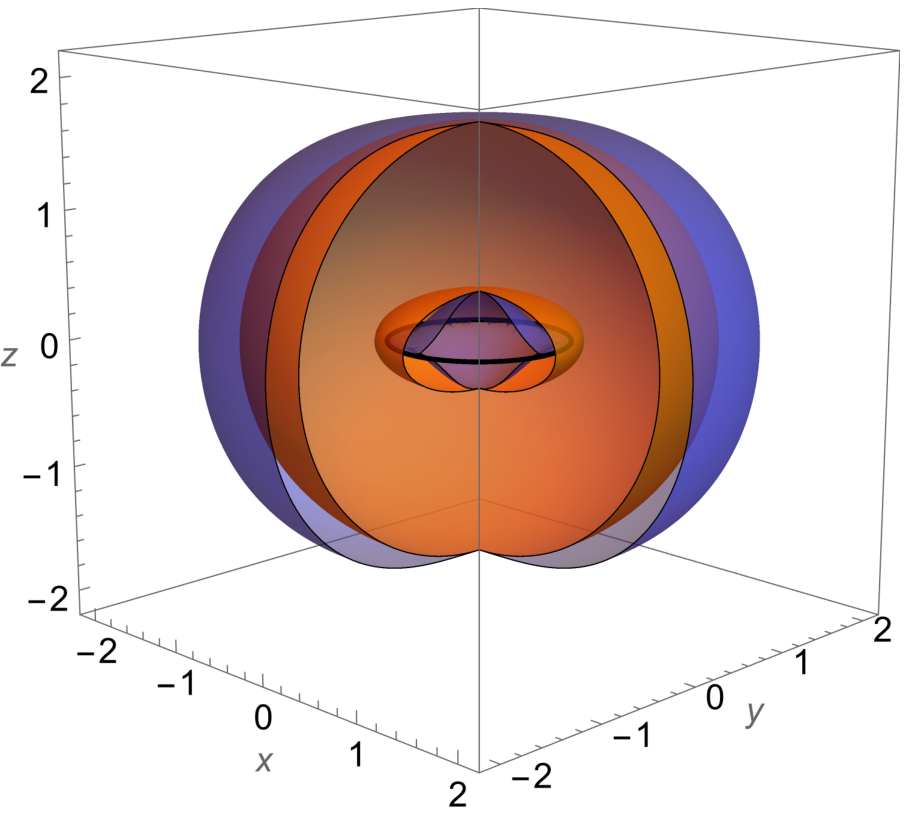}&
		\includegraphics[scale=0.85]{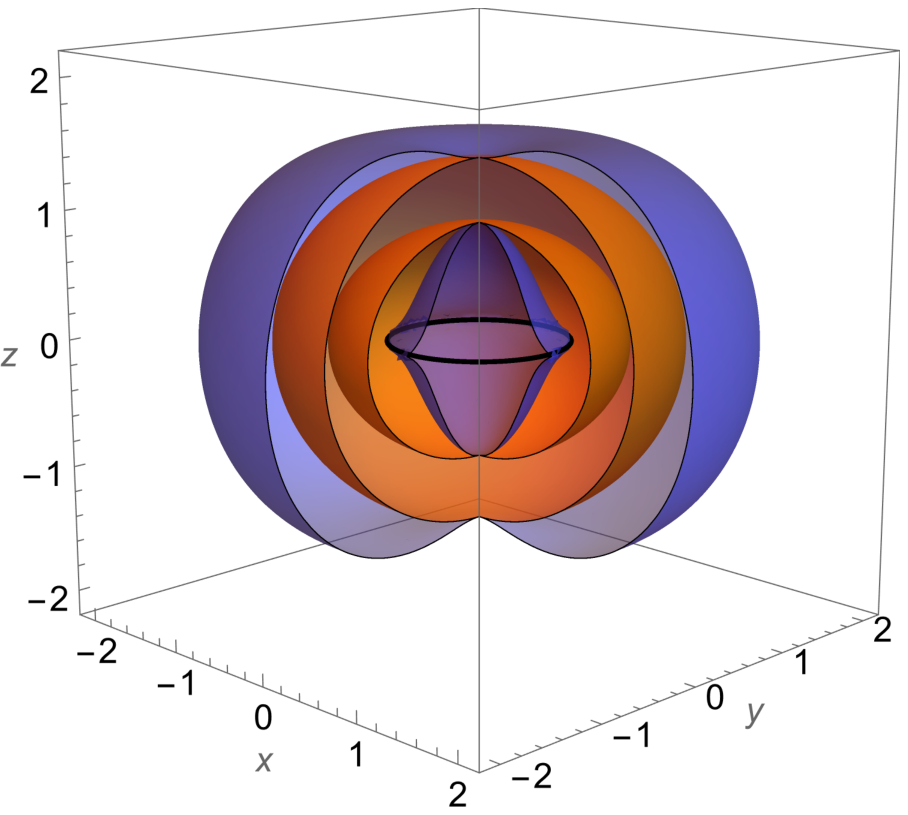}\\
		\includegraphics[scale=0.85]{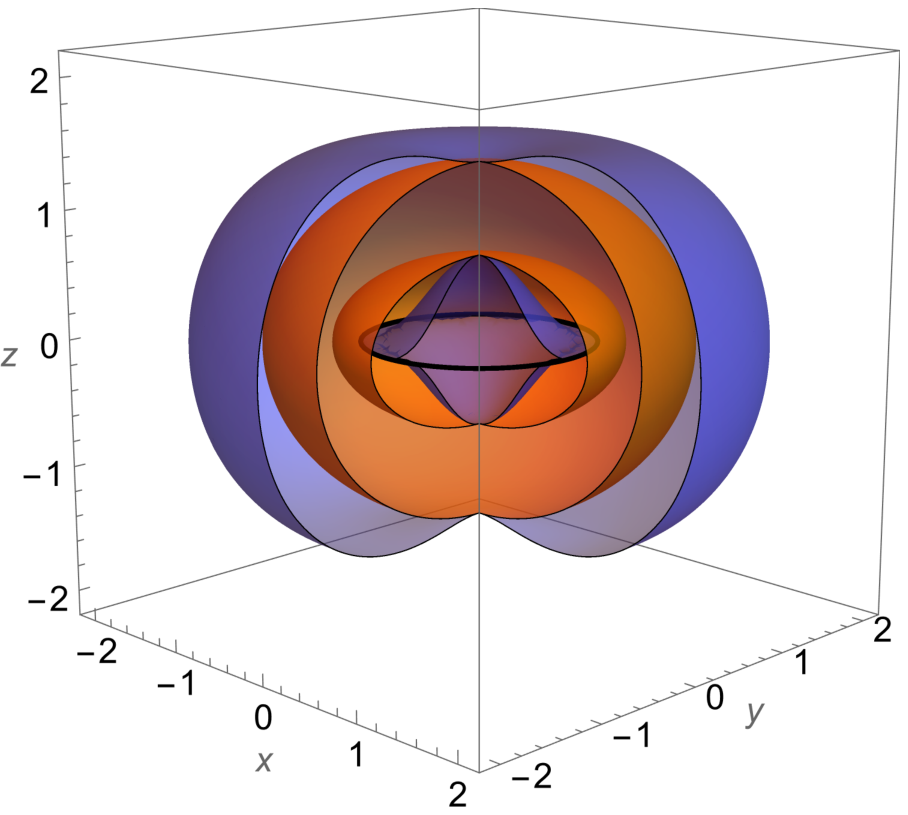}&
		\includegraphics[scale=0.85]{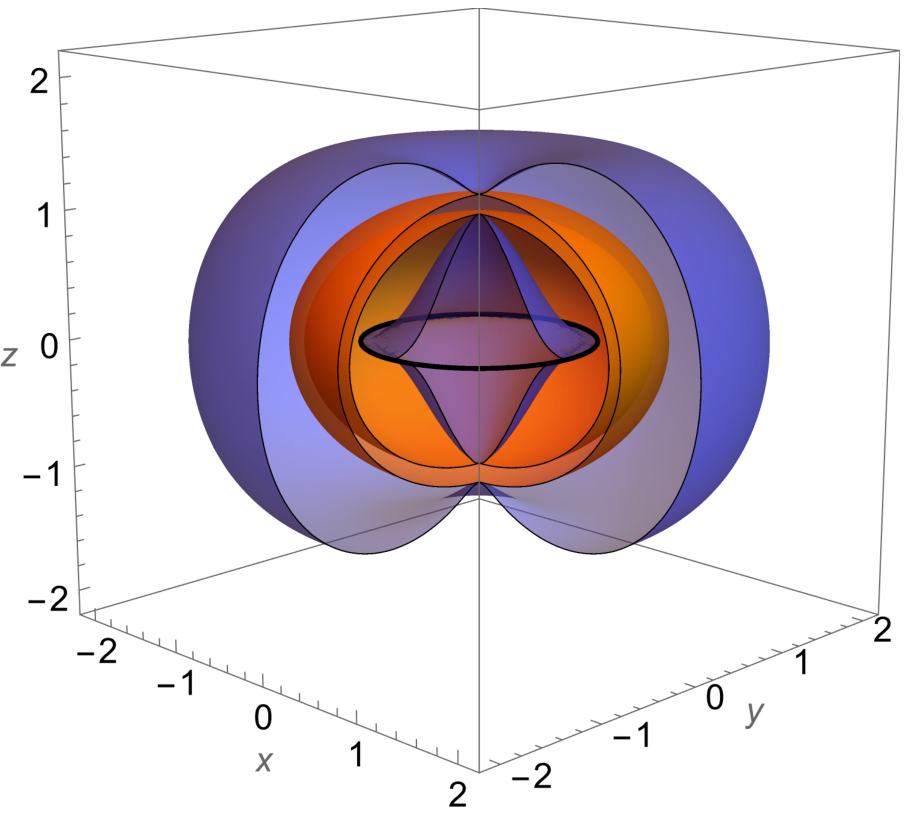}\\
	\end{tabular}
	\caption{Plot showing the ergospheres of rotating hairy black holes. The coordinate transformation from Boyer–Lindquist coordinates ($\lim_{M \rightarrow 0}$) $\{r,\; \theta,\; \phi\}$ to Cartesian coordinates $\{x,\; y,\; z\}$ is given by $x ={\sqrt {r^{2}+a^{2}}}\sin \theta \cos \phi $,\;
		$y ={\sqrt {r^{2}+a^{2}}}\sin \theta \sin \phi \;$,
		$z =r \cos \theta $. Blue surfaces are for the SLS, orange surfaces are for the horizons and black ring is the central singularity. SLS and horizons meet only at $z$-axis. Top right and left: ($a=0.7M$, $Q=-0.2M$) and ($a=0.7M$, $Q=-0.7M$); Bottom right and left: ($a=0.9M$, $Q=-0.1M$) and ($a=0.9M$, $Q=-0.27M$)}\label{plot3}	
\end{figure*}
\begin{figure*}
	\begin{tabular}{c c}
		\includegraphics[scale=0.9]{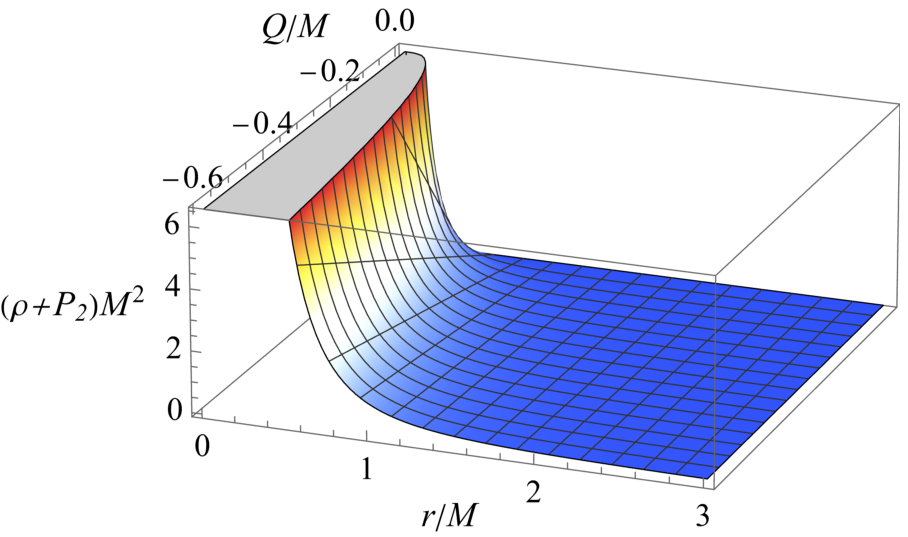} 
		\includegraphics[scale=0.9]{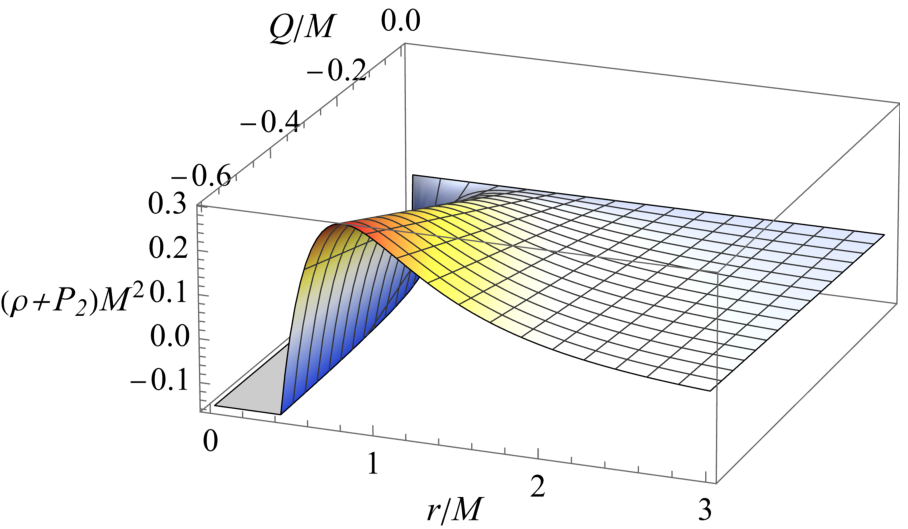} 
	\end{tabular}
	\caption{Plots of $\rho+P_2$ vs radius $r$ and hair $Q$ for $a=0$ (left) and $ a=0.8M $ (right). } \label{rppp}
\end{figure*}
There exist non-vanishing values of parameters $a$ and $Q$ for which $\Delta$ has a minimum, and it admits two positive roots $r_{\pm}$ (with $r_{+} \geq r_{-}$) corresponding to Cauchy ($r_{-}$) and event ($r_{+}$) horizons. Further, analysis of the zeros of $\Delta=0$ reveals, for a given $a$ and $M$, there exists a critical value of $Q$, $Q_E$, such that $\Delta=0$ has two equal roots corresponding to an extremal black hole with degenerate horizons ($r_{-}=r_+=r_E$). When $Q<Q_E$, $\Delta=0$ has two simple zeros, and has no zeros for $Q>Q_E$ (cf. Fig.~\ref{plot2}). They, respectively, correspond to a   non-extremal black hole with a Cauchy horizon and an event horizon, with the a geometry of naked singularity.  

The timelike Killing vector $\xi^a =(\frac{\partial}{\partial t})^a$ of the solution has norm
\begin{equation}
\eta_{(t)}^{\mu} \eta_{(t)\mu}^{}=g_{tt}=-\left(\frac{\Delta-a^2\sin^2\theta}{\Sigma}\right),
\end{equation} 
which is null at the static limit surface (SLS). The timelike Killing vector matches with the time coordinate in the asymptotically large-$r$ limit. The region between $r_+^{H}\, < r\, < r_+^{SLS}$ is called the \textit{ergosphere}, where the asymptotic time translation Killing field $\xi^a=(\frac{\partial}{\partial t})^a$ becomes spacelike and no static observer state is possible. Rotating black holes have two horizons, two SLS, thereby two ergospheres. In Fig.~\ref{plot3}, we depicted rotating Horndeski black hole ergospheres for different values of $a$ and $Q$. With increasing $|Q|$, two horizons come closer and the ergosphere grows. The dependence of ergosphere on the hairy parameter $Q$, in turn, is likely to have an impact on energy extraction by the Penrose process. 

Next, we investigate the stress-energy tensor $T_{\mu\nu}$  and the energy conditions for the source associated with the rotating hairy black hole metric ~(\ref{rotbhtr}). The Kerr and Kerr-Newman black holes belong to a special class of rotating spacetime, which has the same source (trace-free) and thereby the same equation of state both for the rotating and non-rotating cases, i.e., in tetrad frame $|T^{t}_t|=|T^{r}_r|=|T^{\theta}_{\theta}|=|T^{\phi}_{\phi}|$. However, more general rotating black holes require a detailed examination of the correspondence between the stress for the rotating and non-rotating versions. The rotation usually generates momentum density terms in the stress-energy tensor. However, because of the discrete symmetry of the rotating black hole spacetime under joint inversion of $t\to -t$ and $\phi\to -\phi$, the only possible non-zero stress-energy tensor components are $T_{t\phi}$ and $T_{r\theta}$. The stress-energy tensor $T_{\mu\nu}$ in the coordinate basis is cumbersome and contains off-diagonal terms, which can be undone locally with an appropriate
comoving boost. Therefore, we use an orthonormal set of basis in which the stress-energy tensor is diagonal \cite{Bambi:2013ufa,Neves:2014aba,Ghosh:2014pba}
\begin{equation}
e^{(a)}_{\mu}=\left(\begin{array}{cccc}
\sqrt{\mp(g_{tt}-\Omega g_{t\phi})}& 0 & 0 & 0\\
0 & \sqrt{\pm g_{rr}} & 0 & 0\\
0 & 0 & \sqrt{g_{\theta \theta}} & 0\\
{g_{t\phi}}/{\sqrt{g_{\phi\phi}}} & 0 & 0 & \sqrt{g_{\phi\phi}}
\end{array}\right),\label{Matrix}
\end{equation}
with $\Omega= g_{t\phi} /{g_{\phi\phi}}$. The components of the stress-energy tensor in the orthonormal frame reads
\begin{equation}
T^{(a)(b)} = e^{(a)}_{\mu} e^{(b)}_{\nu} G^{\mu \nu}. \nonumber
\end{equation}
Considering the line element (\ref{rotbhtr}), we can write the components of the respective stress-energy tensor as $T^{(a)(b)} = \mbox{diag}(\rho, P_1,P_2,P_3)$ with
\begin{align}
&  \rho = -P_1 = \frac{2r^2M'}{(r^2+a^2)^2} = \frac{-rQ}{(r^2+a^2)^2}
, \nonumber \\
&  P_2 = P_3 = -\frac{r(r^2+a^2)M''+2a^2 M'}{(r^2+a^2)^2} = \frac{Q (a^2-r^2)}{2 r (a^2+r^2)^2},
\end{align}
where, for brevity, we have used $M(r)$ defined in metric (\ref{rotbhtr}). These stresses fall off rapidly at large $r$ for $M \neq 0$ . The weak energy condition requires $\rho\geq0$ (as $Q<0$) and $\rho+P_i\geq0$ ($i=1,\;2,\;3$) \cite{Hawking:1971vc}. Clearly $\rho>0$ and the behaviour of  $ \rho+P_2 = \rho+P_3$ is depicted in Fig.~\ref{rppp}, which shows the weak energy conditions for a  black hole are respected for the spherical case ($a=0$) when $\rho+P_2$ becomes ${-3Q/(2 r^3)}$, but may not be prevented when $a\neq0$ (cf. Fig. \ref{rppp}) for entire $a-Q$ parameter space.  However, the weak energy condition is satisfied for  most region of the parameter space and the violation is minimal, depending on the value of $Q$, as shown in the Fig. \ref{rppp}.  However, this happens for several rotating black holes \cite{Bambi:2013ufa,Ghosh:2014pba, Kumar:2017qws} and, such solutions are important phenomenologically and also they are important from astrophysical point of view being rotating. Furthermore, the nonminimally coupled scalar field models also show the energy condition violation \cite{Brown:2018hym,Klinkhammer:1991ki,Ford:2000xg}.

\section{Komar conserved quantities and Black hole thermodynamics}\label{sec3}
Consider an observer moving with timelike four-velocity in a stationary axisymmetric spacetime; the zero angular momentum observers (ZAMO) are a special class of such observers whose angular momentum vanishes. ZAMO are the stationary observers relative to spatial infinity, but due to frame dragging gain non-zero angular velocity $\omega = {d\phi}/{dt}$. The $\omega$ vanishes for ZAMO at infinity, but in the general case, it is non-zero, and position-dependent \cite{Poisson:2009pwt, Kumar:2017qws}. For the rotating hairy black hole, a stationary observer outside the event horizon, moving with zero angular momentum for an observer at spatial infinity, can rotate with the black hole with an angular velocity given by 
\begin{equation}\label{zamo}
\omega=-\frac{g_{t\phi}}{g_{\phi\phi}}={\frac {a \left( r^2+a^2-\Delta\right) }{ {a}^{4} \sin^{4} \theta -{a}^{2} \left( \Delta-2\,\Sigma \right) \sin^{2} \theta+{\Sigma}^{2}}},
\end{equation}
which at the horizon becomes 	$\omega_+ =a/(r_+^2+a^2)$, i.e., every point of the horizon has the same angular velocity (as measured at infinity). The linear combination of Killing vectors $\eta^{\mu}_{(t)}$ or $\eta^{\mu}_{(\phi)}$ is the generator of the stationary black hole horizon \cite{Chandrasekhar:1992} given by
\begin{equation}
\chi^{\mu}=\eta^{\mu}_{(t)}+\Omega \eta^{\mu}_{(\phi)}.
\end{equation}

The vector $\chi^{\mu}$ is globally time like outside the event horizon, but on the horizon it is a Killing vector \cite{Chandrasekhar:1992}. The Killing vector $\chi^{\mu}=\eta_{(t)}^{\mu}+\Omega\eta_{(\phi)}^{\mu}$ at the event horizon \cite{Kumar:2017qws}, satisfies $\chi^{\mu} \chi_{\mu}=0$, which leads to the rotational velocity 
\begin{equation}
\Omega = \frac{\left[\pm\Sigma\sqrt{\Delta}+a\sin\theta(r^2+a^2-\Delta)\right]}{\left[a^4\sin^4\theta -a^2(-2\Sigma+\Delta)\sin^2\theta+\Sigma^2\right]\sin\theta}.
\end{equation}
Clearly the angular velocity ($\Omega$), even for a fixed $r$, depends on the polar angle $\theta$ and hence $\chi^{\mu}$ is not a Killing vector for arbitrary $r$. 
At horizon $\Delta=0$, we get the black hole rotational velocity 
\begin{equation}\label{frequency}
\Omega_+=\frac{a}{r_+^2+a^2},
\end{equation}
which corresponds to the Kerr black hole value \cite{Poisson:2009pwt,Chandrasekhar:1992}, and at the horizon $\omega_+=\Omega_+$. 
The black hole mass and angular momentum can be identified as the conserved quantities associated with, respectively, $\eta^{\mu}_{(t)} $ and $\eta^{\mu}_{(\phi)}$. Let us consider a spacelike hypersurface $\Sigma_t$, extending from the event horizon to spatial infinity, which is a surface of constant $t$ with unit normal vector $n_{\mu}$ \cite{Chandrasekhar:1992,Wald}. The effective mass reads \cite{Komar:1958wp}
\begin{equation}
M_{\text{eff}}=-\frac{1}{8\pi}\int_{S_t}\nabla^{\mu}\eta^{\nu}_{(t)}dS_{\mu\nu},\label{mass}
\end{equation}
where $S_t$ is the two-boundary of the hypersurface $\Sigma_t$ and is a constant-$t$ and constant-$r$ surface with unit outward normal vector $\sigma_{\mu}$, $dS_{\mu\nu}=-2n_{[\mu}\sigma_{\nu]}\sqrt{h}d^2\theta$ is the surface element of $S_t$, $h$ is the determinant of ($2\times 2$) metric on $S_t$ and 
\begin{equation}
n_{\mu}=-\frac{\delta^{t}_{\mu}}{|g^{tt}|^{1/2}},\qquad \sigma_{\mu}=\frac{\delta^{r}_{\mu}}{|g^{rr}|^{1/2}},
\end{equation}
are, respectively, timelike and spacelike unit outward normal vectors. Thus, the mass integral~(\ref{mass}) becomes an integral over closed 2-surface, that is extendable up to the spatial infinity
\begin{align}
M_{\text{eff}}
=& \frac{1}{4\pi}\int_{0}^{2\phi}\int_{0}^{\phi}\frac{\sqrt{g_{\theta\theta}g_{\phi\phi}}}{|g^{tt}g^{rr}|^{1/2}}\left(g^{tt}\Gamma^{r}_{tt}+g^{t\phi}\Gamma^{r}_{t\phi} \right)d\theta d\phi.
\end{align}
Using the metric (\ref{rotbhtr}), the effective mass of the rotating hairy black hole inside a 2-sphere of radius $r$ is
\begin{align}
M_{\text{eff}}&=\frac{1}{{2 a r}} \Big[{Q \left(a^{2}+r^{2}\right) \arctan \left(\frac{a}{r}\right)+2 \Big(M -\frac{Q }{2} \ln \left(\frac{r}{r_0}\right)\Big) a r}\Big],
\label{mass1}
\end{align}
which is corrected due to the hair parameter $Q$, and in the absence of hair ($Q=0$), reduces to the Kerr black hole mass that is $M_{\text{eff}}=M$. Interestingly, at large distances $r>>r_+$, the effective mass in Horndeski gravity theory is larger than that in  general relativity, i.e., $M_{\text{eff}}>M$. This suggests that gravity is stronger in the Horndeski theory than in general relativity, which leads to interesting observational effects in the gravitational bending of light.
Next, we utilize the spacelike Killing vector $\eta^{\mu}_{(\phi)}$ to calculate the effective angular momentum \cite{Komar:1958wp,Kumar:2017qws}
\begin{equation}
J_{\text{eff}}=\frac{1}{16\pi}\int_{S_t}\nabla^{\mu}\eta^{\nu}_{(\phi)}dS_{\mu\nu},\label{ang}
\end{equation}
using the definitions of surface area element, Eq.~(\ref{ang}) is recast as
\begin{align}
J_{\text{eff}}=&-\frac{1}{8\pi}\int_{0}^{2\phi}\int_{0}^{\phi}\nabla^{\mu}\eta^{\nu}_{(t)}n_{\mu}\sigma_{\nu}\sqrt{h}d\theta d\phi,\nonumber\\
=& \frac{1}{8\pi}\int_{0}^{2\phi}\int_{0}^{\phi}\frac{\sqrt{g_{\theta\theta}g_{\phi\phi}}}{|g^{tt}g^{rr}|^{1/2}}\left(g^{tt}\Gamma^{r}_{t\phi}+g^{t\phi}\Gamma^{r}_{\phi\phi} \right)d\theta d\phi.
\end{align}
Upon using the rotating hairy black hole metric (\ref{rotbhtr}) and integrating, the effective angular momentum becomes 
\begin{align}
J_{\text{eff}}=& \frac {1}{4\,{a}^{2}r} \Bigg[ Q \left( {a}^{2}+{r}^{2} \right) ^{2}
\arctan \left( {\frac {a}{r}} \right)  \nonumber  \\ 
&+4ar\, \left( - \frac{Q{a}^{2}}{2} \ln 
\left( {\frac {r}{r_0}} \right) + \left( M- \frac{Q}{4} \right) {a}^{2}- \frac{Q{r
	}^{2}}{4} \right) \Bigg]. \label{ang1}
\end{align}
In the absence of hair, $Q \to 0$, the effective angular momentum Eq.~(\ref{ang1}) gives the Kerr black hole angular momentum value $J_{\text{eff}}=Ma$.
The Killing vector $\chi^{\mu}= \eta^{\mu}_{(t)}+\Omega \eta^{\mu}_{(\phi)}$ is the generator of the rotating black hole Killing horizon, the corresponding Komar conserved quantity reads as \cite{Komar:1958wp}
\begin{eqnarray}
\mathcal{K}_{\chi}&=&-\frac{1}{8\pi}\int_{S_t}\nabla^{\mu}\chi^{\nu}dS_{\mu\nu}, \nonumber\\
&=&-\frac{1}{8\pi}\int_{S_t}\nabla^{\mu}\left( \eta^{\mu}_{(t)}+\Omega \eta^{\mu}_{(\phi)}\right)dS_{\mu\nu}.
\end{eqnarray}
Using Eqs.~(\ref{mass1}) and (\ref{ang1}), we obtain
\begin{eqnarray}
\mathcal{K}_{\chi}&=& \frac{1}{2 r_{+}} \left[{r_+^2+Q r_{+} -a^{2}}\right]
.\label{ST}\end{eqnarray} 

\begin{figure*}
	\begin{tabular}{c c}
		\includegraphics[scale=0.85]{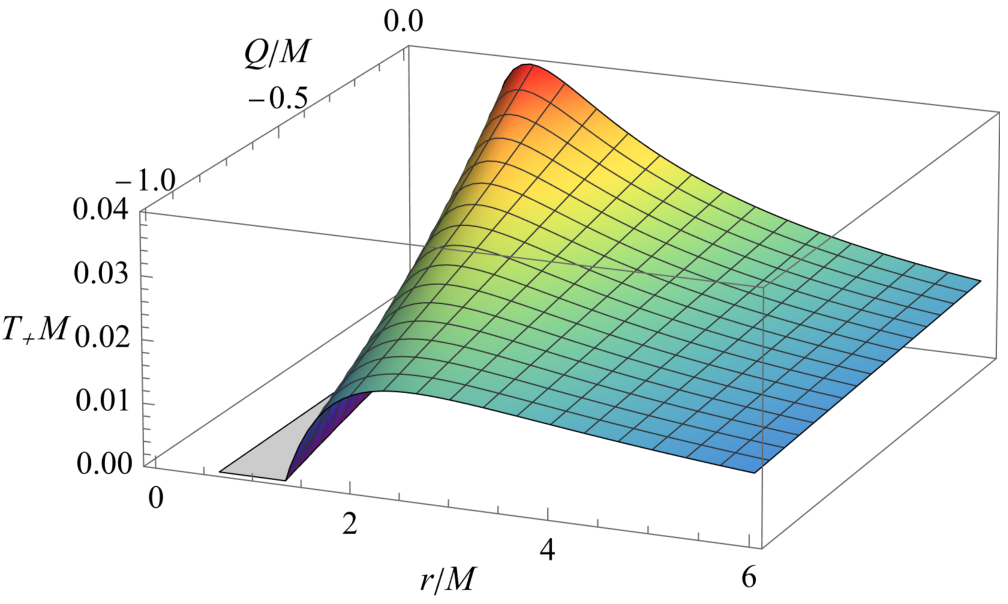} 
		\includegraphics[scale=0.85]{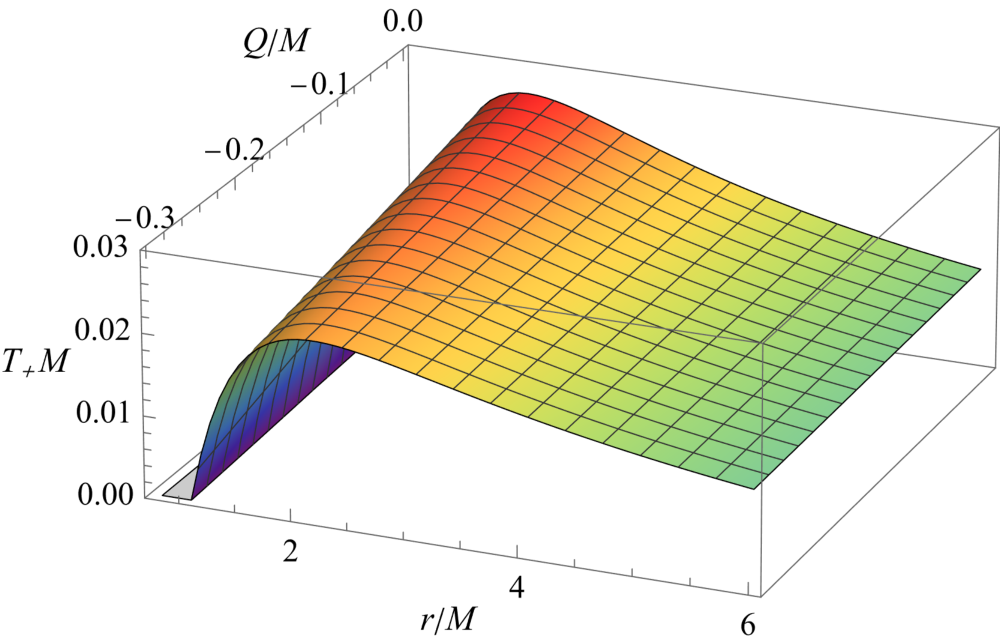} 
	\end{tabular}
	\caption{ Plot of temperature $T_+$ vs horizon radius $r_+$ for $a=0.6M$ and $0.9M.$} \label{ptemp}	
\end{figure*}

Next, we analyse the thermodynamical quantities associated with the rotating hairy black hole metric (\ref{rotbhtr}). The surface gravity $\kappa$, the angular velocity $\Omega$ and the electrostatic potential $\Phi$ are all locally defined on the horizon and they are always constant over the horizon of any stationary black hole, leading to an extended form of the zeroth law of black hole mechanics. The surface area of the black hole horizon is give by \cite{Poisson:2009pwt,Kumar:2017qws}
\begin{equation}
A_H=\int_{0}^{2\pi}d\phi\int_{0}^{\pi}\sqrt{g_{\theta\theta}g_{\phi\phi}}d\theta,
\end{equation}
which upon integration leads to 
$ A_H=4\pi(r_+^2+a^2)$, $r_+$ is outer horizon radius. The entropy of the black hole can be calculated using Hawking's area law of black hole as 
\begin{equation}\label{entropy}
S_{+}=\frac{A_H}{4}=\pi (r_+^2+a^2).
\end{equation}
The black holes are characterized by their mass $(M_+)$. 
From $\Delta (r_+)=0$, the mass of the black hole can be expressed in terms of its horizon radius 
($r_+$) by
\begin{equation}\label{BHMass}
M_+ = \frac{1}{{2 r_+}} \left[r_+^2+a^2+r_+ Q \ln \! \left(\frac{r_+}{r_0}\right)\right],
\end{equation}
which in the special case $Q=0$, reduces to $M_{+}= ({{a}^{2}+r_{+}^{2}})/{(2\,r_+)}$, corresponding to the Kerr black hole mass \cite{Poisson:2009pwt}. Black holes behave as a thermodynamical entity whose temperature $T$ can be calculated from surface gravity $\kappa$ evaluated at Killing horizon through $\kappa^2=-\frac{1}{2}\chi^{\mu;\nu}\chi_{\mu;\nu}.$ Hawking showed that the black hole temperature is determined by
\begin{eqnarray}
T_+= && \frac{\kappa}{2\pi}=\frac{\Delta'(r_+)}{4\pi(r_+^2+a^2)} \nonumber \\
= && {\frac {r_+^2-a^2+Qr_{+}}{4\pi \,r_+ \left( r_+^2+a^2 \right) }},	\label{temp}
\end{eqnarray}
which in the absence of hair ($ Q=0 $), yields
\begin{eqnarray}
T_+^{K}&=& {\frac {r_{+}^{2} - {a}^{2}}{4 \pi\,r_+ \left(r_+^2+a^2 \right) }}, 
\label{tempK}
\end{eqnarray}
the temperature of the Kerr black hole \cite{Kumar:2017qws,Kumar:2020hgm}. It is evident from Fig. \ref{ptemp} that, with decreasing $r_+$, the Hawking temperature of the rotating hairy black holes increases to a maximum $T^{max}_+$ at the critical radius $r_c$, then drops to zero temperature and then becomes negative at a small horizon radius. The maximum Hawking temperature $T^{max}_+$ decreases with an increase in the values of the hairy parameter $Q$. The Hawking temperature of the rotating hairy black hole spacetime increases with the horizon radius (or with the mass) (cf. Fig. \ref{ptemp} ) when horizon radii in the range $r_E \leq r_+ \leq r_C$. 

\begin{figure*}
	\begin{tabular}{c c}
		\includegraphics[scale=0.85]{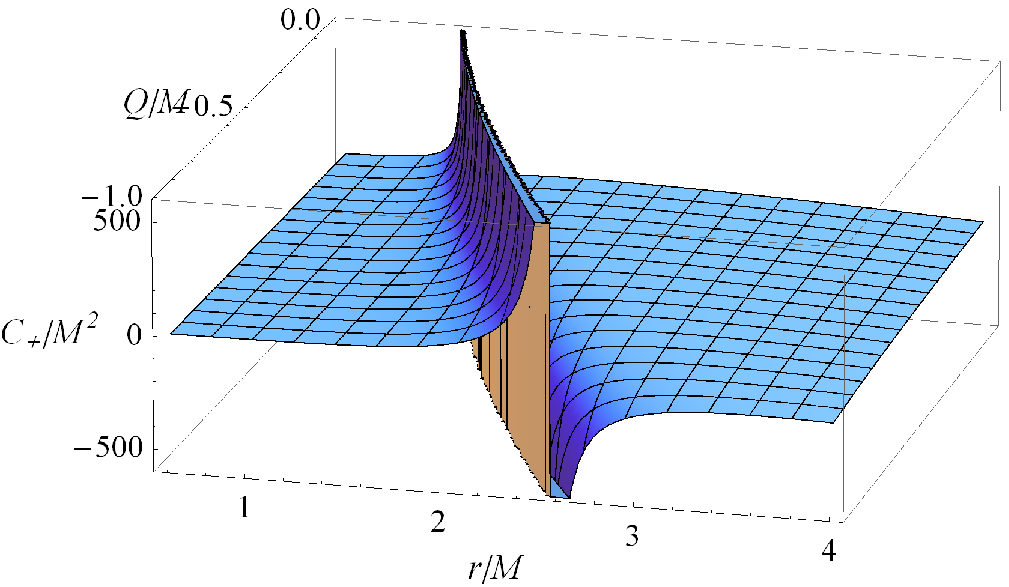} \hspace*{0.2cm}&
		\includegraphics[scale=0.85]{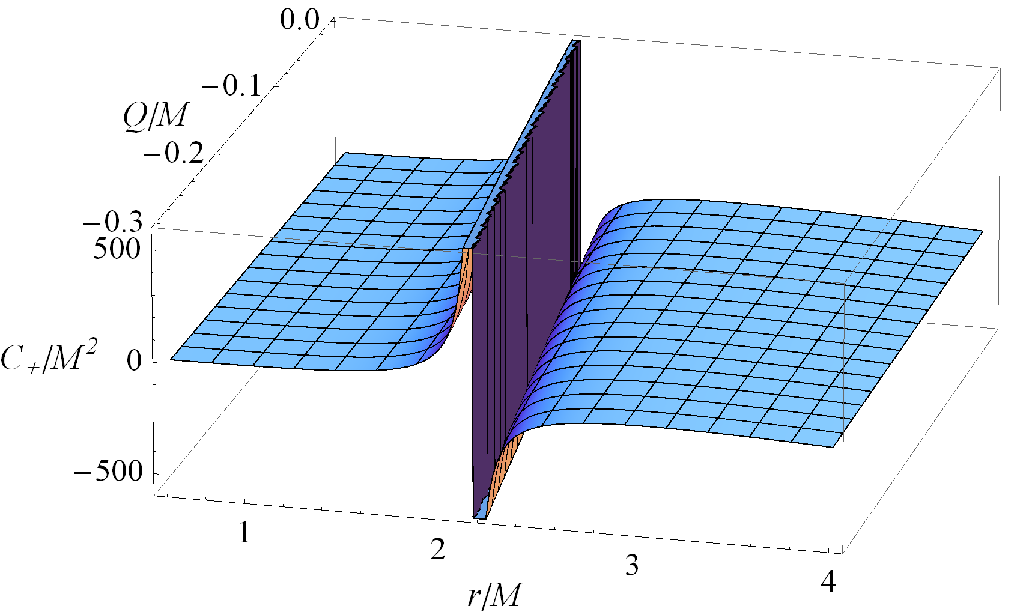} 
	\end{tabular}
	\caption{Plots of $C_{+}$ vs radius $r_+$ and hair $Q$ for $a=0.6M$ and $0.9M.$ }\label{stab}
\end{figure*}

\begin{figure*}
	\begin{tabular}{c c}
		\includegraphics[scale=0.85]{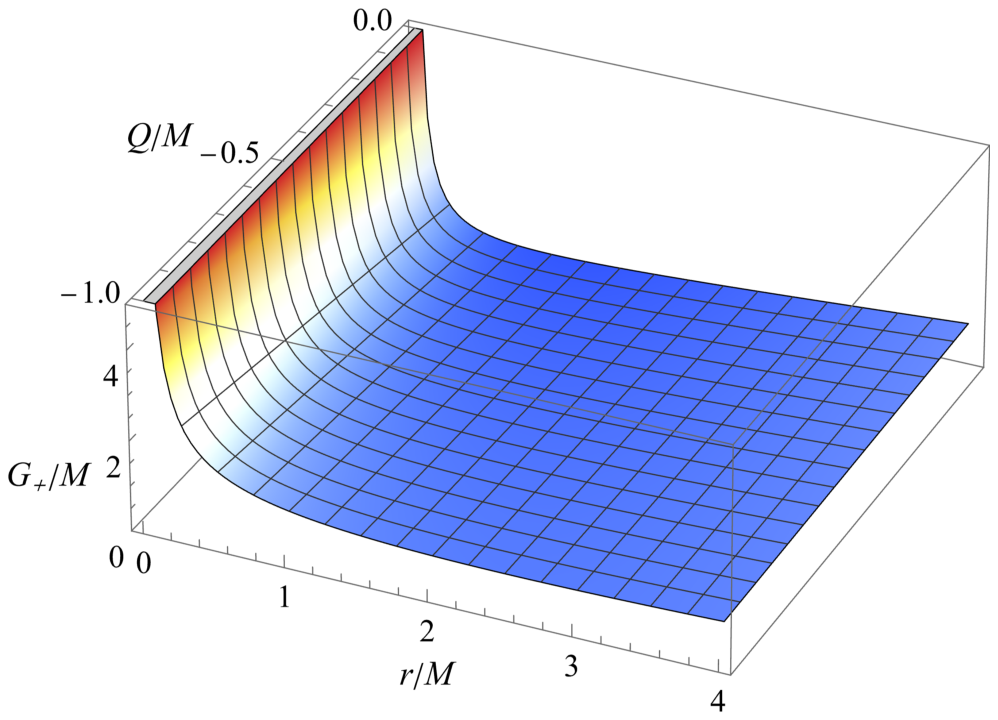} \hspace*{0.2cm}&
		\includegraphics[scale=0.85]{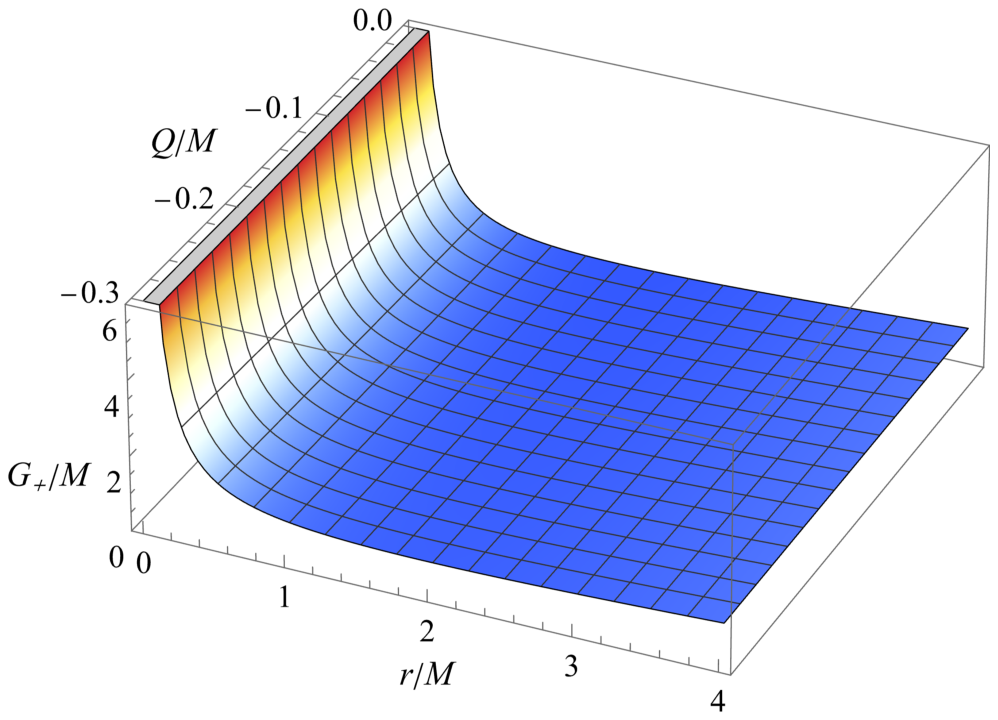}  
	\end{tabular}
	\caption{Plots of $G_{+}$ vs radius $r_+$ and hair $Q$ for $a=0.6M$ and $a=0.9M$. }\label{stab1}
\end{figure*}

From classical electrodynamics we can calculate the potential $ \Phi$ associated with black hole charge $Q$
\cite{Chen:2008ra, Sekiwa:2006qj}. Because the rotating Horndeski gravity black hole is asymptotically flat at $r\to\infty$, therefore, the differential form of the first law of black hole thermodynamics can be written as
\begin{equation}\label{flaw}	
dM=TdS+\Omega dJ+\Phi dQ.
\end{equation}
Furthermore, using this we can calculate the extensive quantity associated with black hole, i.e., temperature, angular velocity and electrostatic potential, through
\begin{equation}\label{first}
T=\left(\frac{dM}{dS}\right)_{(J,Q)},\;
\Omega=\left(\frac{dM}{dJ}\right)_{(S,Q)},\;
\Phi=\left(\frac{dM}{dQ}\right)_{(S,J)}.
\end{equation}
Lastly, Eqs (\ref{ST}), (\ref{temp}) and (\ref{entropy}), lead to 
\begin{equation}
\mathcal{K}_{\chi}=M_{\text{eff}}-2\Omega J_{\text{eff}}=2S_+T_+.\label{smarr}
\end{equation}
Therefore, the Komar conserved quantity corresponding to the null Killing vector at the event horizon $\chi^{\mu}$ is twice the product of the black hole entropy and the horizon temperature and hence satisfies the Smarr formula \cite{Smarr:1972kt,Bardeen:1973gs,Kumar:2017qws}. It is worthwhile to notice that Eq.~(\ref{flaw}) is in terms of ADM quantities whereas Eq.~(\ref{smarr}) is in terms of the Komar quantity, which does not have an explicit contribution from the charge $Q$. 

Finally, we analyse the thermodynamic stability of the rotating  black hole, which requires the study of its heat capacity that is defined as \cite{Cai:2003kt,Sahabandu:2005ma}
\begin{equation}\label{sh_formula}
C_+ = \frac{\partial{M_+}}{\partial{T_+}}= \left(\frac{\partial{M_+}}{\partial{r_+}}\right)
\left(\frac{\partial{T_+}}{\partial{r_+}}\right)^{-1}. 
\end{equation} 
The global thermodynamical stability of the black hole can be deduced from the behaviour of its free energy. The Gibb's free energy of the black hole in the canonical ensemble is obtained as \cite{Altamirano:2014tva,Carlip:2003ne}
\begin{eqnarray}\label{fe}
G_+&=&M_+-T_+S_+.
\end{eqnarray} 
On using Eqs.~(\ref{entropy}), (\ref{BHMass}) and (\ref{temp}) in Eqs. (\ref{sh_formula}) and (\ref{fe}), we get expressions, respectively, for heat capacity and Gibb's free energy. The graphic results, depicted in Fig. \ref{stab}, might be more enlightening than the analytical expressions given by
\begin{eqnarray*}
	C_{+} & = &-\frac{2\pi\left(r_+ Q -a^{2}+r_+^{2}\right) \left(r_+^2+a^2\right)^{2}}{2 Q \,r_+^{3}-a^{4}-4 a^{2} r_+^{2}+r_+^{4}}, \nonumber \\
	G_{+} & = &	\frac{1}{{4 r_+}} \left[r_+^2+3a^2+2r_+ Q \ln \left(\frac{r_+}{r_0}\right)-r_+Q\right].
\end{eqnarray*}
However, as a consistency check, the expression for heat capacity and Gibb's free energy for the Kerr black hole recovered when $Q=0$ \cite{Kumar:2017qws,Kumar:2020hgm,Altamirano:2014tva}. 
The positivity of black hole's heat capacity $C_+>0$ is sufficient to state that the black hole is thermodynamically stable to local thermal fluctuation; black hole temperature increases with its mass. When the specific heat is positive, an increase in the black hole temperature will increase the entropy, thereby giving a stable thermodynamic configuration. Fig. \ref{stab} shows that heat capacity, for a given value of $Q$ and $a$, is discontinuous at a critical radius $r^{C}_+$. Further, we notice that the heat capacity flips its sign around $r^{C}_+$. Thus, we can say the black hole is thermodynamically stable for $r_1<r_+ < r^{C}_+$ where ${C}_+>0$, whereas it is thermodynamically unstable for $r_+>r_+^C$ region wherein ${C}_+<0$, and there is a second order phase transition at $r_+=r^{C}_+$.  Thus, we can say that the heat capacity is negative for a larger black hole with $r_+>r_+^C$, positive for the black hole in the region $r_1<r+<r_+^C$ and again negative for minimal radius $r<r_1$. It means that the smaller size black holes are thermodynamically stable locally \cite{Cai:2003kt,Sahabandu:2005ma,Ghosh:2014pga}. One can notice that the black hole temperature decreases with the increasing $r_+$ for $r_+>r_c$, thus leaving a thermodynamically unstable black hole with $C_+<0$. Whereas at the critical radius $r_c$, the black hole temperature is maximum and the specific heat is discontinuous. 

One can analyze the free energy to discuss the global thermodynamical stability of black holes \cite{Altamirano:2014tva,Kumar:2017qws}. If we consider that the black hole is in thermal equilibrium with the surrounding radiation, such that it exchanges only energy, then in the preferred phase, free energy will be minimum. A black hole, at some stage, due to Hawking evaporation, absorbs more radiation than it emits, which leads to global stability \cite{Cai:2003kt,Sahabandu:2005ma, Ghosh:2014pga}. As we all know, the thermodynamic state with lower Gibbs free energy is more stable. We depict the Gibbs free energy for various values of parameter $Q$ and $a$ in Fig.~\ref{stab1}. We note that the free energy has a positive value for the entire parameter space $(r_+,\;Q)$. Thus rotating Horndeski gravity black holes are globally thermodynamically unstable. However, the black holes with large $r_+$ have smaller Gibbs free energy. The Hawking-Page-type phase transition is not possible as the free $G_{+}>0$ for all $r_+$ as depicted in Fig.~\ref{stab1}. 

\section{Gravitational deflection of light}\label{sec4}
It is a well established fact now that the light rays get deflected when they propagate through an inhomogeneous gravitational field, such as near the black hole, and the phenomenon is known as gravitational lensing. The gravitational lensing has been used as an important tool to make a precision test of theories of gravity and to detect exotic objects in the universe \cite{Virbhadra:2008ws,Virbhadra:2007kw,Virbhadra:2002ju}. Here, we focus on weak gravitational lensing by the rotating Horndeski black hole for a source, and observer at a finite distance from the black hole. In particular, to calculate the deflection angle, we use the Gauss-Bonnet theorem that connects the differential geometry of the surface of light propagation with its topology or commonly known as the Gibbons-Werner method \cite{Gibbons:2008rj}. In the geometrical approach to gravitational lensing theory, Gibbon and Werner \cite{Gibbons:2008rj} devised an elegant technique to calculate the deflection angle of light in a static and spherically symmetric black hole spacetime in the context of optical geometry. They reported that the focusing of light rays emerges as a global topological effect, and consequently, the deflection angle can be calculated by integrating the Gaussian curvature of the optical metric outwards from the light ray \cite{carmo}. Their method was extended to take account of the finite distance from a black hole to a light source and a receiver by Ishihara \textit{et al.} \cite{Ishihara:2016vdc}, who calculated the light deflection angle for static and spherically symmetric black holes (including the presence of the cosmological constant). Later, Ono, Ishihara, and Asada \cite{Ono:2017pie} used this method to calculate the finite-distance correction for stationary and axisymmetric spacetimes. For both finite or infinite distant sources and observers, the Gauss-Bonnet theorem is often applied to an infinite region outside the light ray. This method has gained popularity over time and has been extensively used for varieties of black hole spacetimes \cite{Crisnejo1:2018uyn}. 

We begin by determining the photon orbits equation of motion at the equatorial plane ($\theta=\pi/2$) for the rotating Horndeski gravity black hole. Because the rotating metric is isometric under time translation and rotational transformation along the $\phi$ axis, this yields two constants of motion for photons, namely energy $E$ and axial angular momentum $L$
\begin{align}
E=& \left(1-\frac{2 m(r)}{r}\right)\dot{t} + \frac{2m(r)a}{r} \dot{\phi},\nonumber\\
L=& -\frac{2m(r)a}{r}\dot{t}+\left(r^2+a^2+\frac{2m(r)a^2}{r}\right) \dot{\phi},
\end{align}
where `` $\dot{}$ " denotes the derivative with respect to the affine parameter along geodesics. The impact parameter $b$ for the photon trajectories is defined as 
\begin{equation}
b=\frac{L}{E}.
\end{equation}
In terms of the impact parameter the light orbit equation ($ds^2=0$) takes the following simpler form
\begin{align}
\Big(\frac{d\,u}{d\,\phi}\Big)^2&=\Big(1+a^2u^2-2Mu+Qu\ln[\frac{1}{u\,r_0}]\Big)^2\Big(1+\nonumber\\
&(a^2-b^2)u^2+(a-b)^2u^3(2M-Q\ln[\frac{1}{u\,r_0}])\Big)\nonumber\\
&\times \frac{1}{\Big(b+(a-b)u(2M-Q\ln[\frac{1}{u\,r_0}])\Big)^2}\equiv F[u] ,\label{Orbit}
\end{align}
where $u=1/r$.
\begin{figure*}
	\begin{tabular}{c c}
		\includegraphics[scale=0.78]{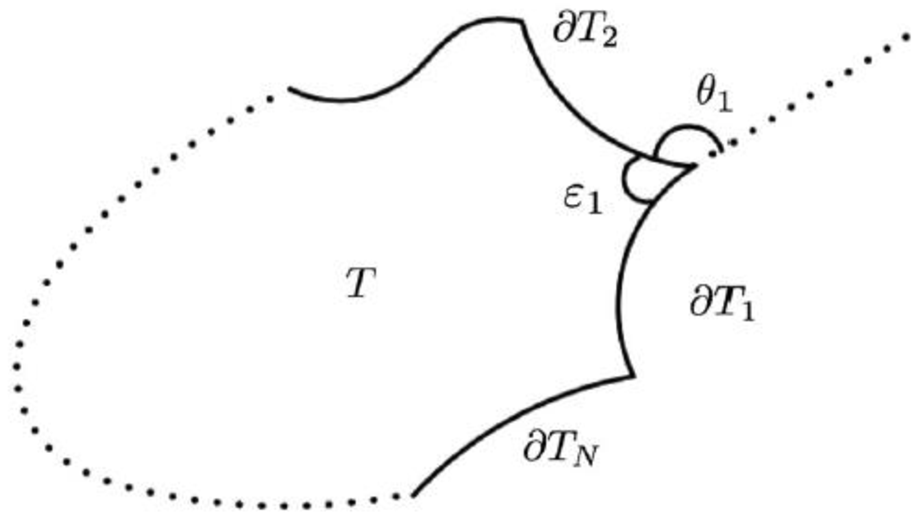} \hspace*{0.2cm}&
		\includegraphics[scale=0.60]{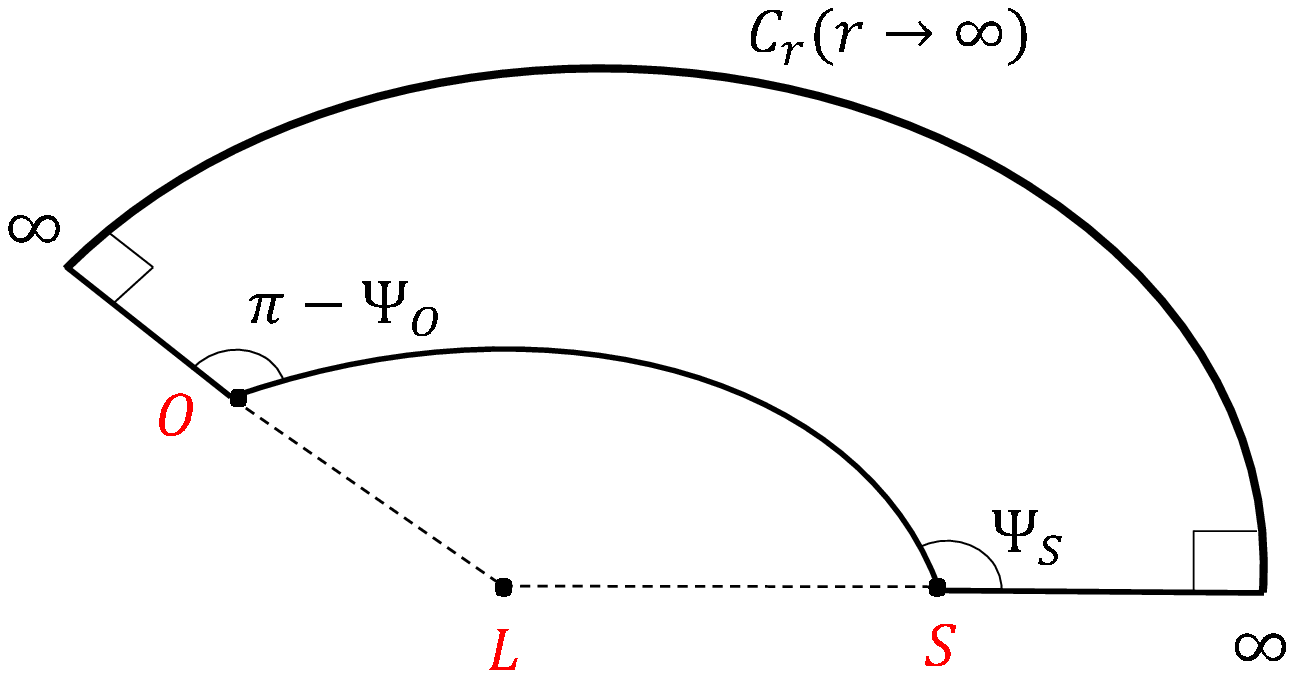}  
	\end{tabular}
	\caption{\textbf{Left:} A schematic for the Gauss-Bonnet theorem \cite{Gibbons:2008rj}. \textbf{Right:} Geometrical setup for weak-gravitational lensing. Note that the inner angle at the source and observer are $\Psi_S$ and $\pi-\Psi_O$.}\label{lensing1}
\end{figure*}

For a compact and oriented two-dimensional surface $T$ with boundary curves $\partial T_p$, such that the \textit{exterior} jump angle between the curves $\partial T_p$ and $\partial T_{p+1}$ along the boundary is $\theta_p$, the Gauss-Bonnet theorem can be mathematically expressed as \cite{Gibbons:2008rj}
\begin{equation}
\int\int_T K dS+\Sigma_{p=1}^{N}
\int_{\partial T} k_g d\ell +\Sigma_{p=1}^{N} \theta_p=2\pi\chi(T),\label{GBEq}
\end{equation}
where $K$ is the Gaussian curvature at any point on the surface, $dS$ and $d\ell$ are the infinitesimal surface area element and line element, $k_g$ is the curvature of boundary curve $T_p$ and $\chi(T)$ is the Euler characteristic of the surface. For the gravitational lensing setup, by construction, both the source ($S$) and observer ($O$) are considered at a finite distance from the black hole ($L$) (cf. Fig.~\ref{lensing1}), and the deflection angle of light is defined as
\cite{Ono:2017pie}
\begin{equation}
\alpha_D=\Psi_O-\Psi_S+\Phi_{OS}.\label{defEq}
\end{equation}
Here, $\Psi_S$ and $\Psi_O$, respectively, are angles made by light rays tangent and the radial direction from the lens object at the source and observer, and $\Phi_{OS}$ is the source-observer angular coordinate separation that is associated with the rotational Killing vector $\phi$ (cf. Fig.~\ref{lensing1}). On using Eq.~(\ref{Orbit})
\begin{equation}
\alpha_D=\Psi_O-\Psi_S+\int_{u_o}^{u_m} \frac{du}{\sqrt{F(u)}}+\int_{u_s}^{u_m} \frac{du}{\sqrt{F(u)}},\label{defEq1}
\end{equation}
where $u_{s,o}$ are the inverse of the distance of source and observer from the black hole $r_{s,o}$ and $u_m$ is the inverse of distance of closest approach to the black hole $r_m$. In the far limit of source and observer, i.e., $u_{o,s}=0$, $\Psi_O\to 0$ and $\Psi_s\to \pi$, Eq.~(\ref{defEq1}) yields
\begin{equation}
\alpha_D=2\int_{0}^{u_m} \frac{du}{\sqrt{F(u)}}-\pi.\label{defEq2}
\end{equation}
It comes as a great surprise that Eqs.~(\ref{defEq}) and (\ref{defEq1}) are valid for any winding number of photon orbits around the black hole \cite{Ishihara:2016sfv}.
To use the Gauss-Bonnet theorem \cite{Gibbons:2008rj}, we take into account a domain ${}_O^{\infty}\Box_{S}^{\infty}$ constructed by a spatial light ray curve from $S$ to $O$, two 
outward straight lines from $O$ and $S$ to $O_{\infty}$ and $S_{\infty}$, and a circular arc segment $C_r$ of coordinate radius $r_C$ $(r_C\to\infty)$ (cf. Fig.~\ref{lensing1}). The outward radial lines $LO_{\infty}$ and $LS_{\infty}$ meets the circular arc at the right angle. This domain ${}_O^{\infty}\Box_{S}^{\infty}$ is defined by a three-dimensional spatial metric $\sigma_{ij}$, in which photon trajectories are described as spatial curves \cite{Gibbons:2008rj}. To determine the spatial metric, we solve the null geodesics equation $ds^2=0$ in four-dimensional spacetime, this reads as
\begin{equation}
dt= \pm\sqrt{\sigma_{ij}dx^i dx^j}+\beta_i dx^i,
\end{equation}
with $i,j=1,2,3$ and
\begin{align}
\sigma_{ij}dx^i dx^j=&\frac{\Sigma^2}{\Delta(\Delta-a^2\sin^2\theta)}dr^2+\frac{\Sigma^2}{\Delta-a^2\sin^2\theta}d\theta^2\nonumber\\
&+ \left(r^2+a^2+\frac{2m(r)ra^2\sin^2\theta}{\Delta-a^2\sin^2\theta}\right)\frac{\Sigma\sin^2\theta\, d\phi^2}{(\Delta-a^2\sin^2\theta)} ,\nonumber\\
\beta_idx^i=&-\frac{2m(r)ar\sin^2\theta}{\Delta-a^2\sin^2\theta}d\phi.\label{metric3}
\end{align}
The advantage of the spatial metric $\sigma_{ij}$ is that the null arc length $d\ell$ is directly related with the time associated with the timelike Killing vector. At the equatorial plane, the two-dimensional metric $\sigma_{ij}^{(2)}$ defines the surface of light propagation with Gaussian curvature $K$ and infinitesimal area element $dS$. The arc-length along the null ray at $\theta=\pi/2$ reads as
\begin{equation}
d\ell^2=\frac{r^4}{\Delta(\Delta-a^2)}dr^2+ \left(r^2+a^2+\frac{2m(r)ra^2}{\Delta-a^2}\right)\frac{\Sigma\, d\phi^2}{(\Delta-a^2)}.\label{lenght}
\end{equation} 
Because of the gravitomagnetic effect in rotating black holes, the null geodesics of $g_{\mu\nu}$ are no longer the geodesics of $\sigma_{ij}$ and rather are just the spatial curves. 

As depicted in Fig.~\ref{lensing1} the two outgoing radial lines $LO_{\infty}$ and $LS_{\infty}$ have zero geodesic curvature. Furthermore, in the asymptotically flat spacetime, the curvature of circular arc $C_r$ is $k_g=1/r_c$ with $d\ell=r_c\,d\,\phi$, this yields
\begin{equation}
\int_{C_r} k_g\,d\ell=\int_{C_r} d\phi= \Phi_{OS}.
\end{equation} 
For the domain ${}_O^{\infty}\Box_{S}^{\infty}$ shown in Fig.~\ref{lensing1} and $\chi(T)=1$, the Gauss-Bonnet equation~(\ref{GBEq}) can be simplified as
\begin{align}
&\int\int_T K dS+ \int_{O}^{S} k_g d\ell+\Phi_{OS} +\frac{\pi}{2}+\frac{\pi}{2}+\Psi_O\nonumber\\
&\quad +(\pi-\Psi_S)=2\pi,\nonumber\\
\Rightarrow &\int\int_T K dS+ \int_{O}^{S} k_g d\ell+\Psi_O-\Psi_S+\Phi_{OS}=0,
\label{GBEq1}
\end{align}
which on using Eq.~(\ref{defEq}) yields the geometrically invariant definition of deflection angle as follows \cite{Ono:2017pie}
\begin{equation}
\alpha_D=-\int\int_{{}_O^{\infty}\Box_{S}^{\infty}} K dS+\int_{S}^{O} k_g d\ell.\label{deflectionangle}
\end{equation}
Here, $k_g$ denotes the geodesic curvature of light curves from $S$ to $O$, which is the surface-tangential component of the acceleration of the curve \cite{Ono:2017pie} and reads as
\begin{equation}
k_g=-\frac{1}{\sqrt{\sigma\sigma^{\theta\theta}}}\beta_{\phi,r}.
\end{equation}

Therefore, for a stationary spacetime, both the Gaussian curvature of the surface of light propagation defined by spatial metric $\sigma_{ij}$ and the geodesics curvature of light curves contribute to the light deflection angle $\alpha_D$. However, for a static spacetime, the geodesic curvature contribution vanishes. We calculate both contributions. We begin with the Gaussian curvature of the two-dimensional surface, which is defined as \cite{Werner:2012rc}
\begin{align}
K=&\frac{{}^{3}R_{r\phi r\phi}}{\sigma}\nonumber\\
=&\frac{1}{\sqrt{\sigma}}\left(\frac{\partial}{\partial \phi}\left(\frac{\sqrt{\sigma}}{\sigma_{rr}}\Gamma^{\phi}_{rr}\right) - \frac{\partial}{\partial r}\left(\frac{\sqrt{\sigma}}{\sigma_{rr}}\Gamma^{\phi}_{r\phi}\right)\right),
\end{align}
where $\sigma=\det(\sigma_{ij})$. For the rotating Horndeski black hole, this reads as
\begin{align}
K&=\frac{-1}{6r^5(r-2m(r))}\Big(6r^2\left(r-2m(r)\right)(\Delta+a^2)m''(r)\nonumber\\
&+6rm'(r)(rm'(r)-m(r))(\Delta+5a^2)+6rm(r)^2\nonumber\\
&-(7r^2+a^2)m(r)+2r(r^2+3a^2) \Big),\label{K}
\end{align}
on using the black hole mass function and using the weak-field approximation, this reduces to
\begin{align}
K=&-\frac{3Q}{2r^3}-\frac{Q^2}{4r^4} -\frac{4a^2Q}{r^5}+\frac{Q\ln(r/r_0)}{r^3}-\frac{Q^2\ln(r/r_0)}{ r^4}\nonumber\\
&- \Big(\frac{2}{r^3}+\frac{3Q\ln(r/r_0)}{r^4}-\frac{2Q}{r^4}+\frac{6a^2}{r^5} +\frac{6a^2Q}{r^6}\nonumber\\
&-\frac{6a^2Q\ln(r/r_0)}{r^6}\Big)M +\Big(\frac{3}{r^4}-\frac{6a^2}{r^6}\Big)M^2\nonumber\\
&	+\mathcal{O}\Big(\frac{Q^2a^2}{r^6},\frac{MQ^2a^2}{r^7}\Big).
\end{align}
This captures only the leading order contributing terms. To perform the surface integral of Gaussian curvature over the closed quadrilateral ${}_O^{\infty}\Box_{S}^{\infty}$, we have to identify the boundary of the integration domain, i.e., the curve from $S$ to $O$. In the weak field approximation, we can use the light orbit solution from Eq.~(\ref{Orbit}) \cite{Crisnejo:2019ril}:
\begin{equation}
u=\frac{\sin\phi}{b} + \frac{M(1-\cos\phi)^2}{b^2} -\frac{2Ma(1-\cos\phi)}{b^3}+\mathcal{O}\Big(\frac{M^2}{b^3}\Big)\label{orbit1}
\end{equation}
The surface integral yields
\begin{align}
\int\int_{{}_O^{\infty}\Box_{S}^{\infty}} K dS=& \int_{\phi_S}^{\phi_O}\int_{\infty}^{r} K \sqrt{\sigma}dr d\phi \nonumber\\
& =\int_{\phi_S}^{\phi_O}\int_{0}^{u}-\frac{K\sqrt{\sigma}}{u^2}du d\phi.\label{Gaussian}
\end{align}

On evaluating Eq.~(\ref{Gaussian}), we get

\begin{align}
&\int\int K dS=\int_{\phi_S}^{\phi_O}\int_{0}^{u} \frac{-K}{u^2}du\,d\phi\nonumber\\
=&-\frac{1}{2b}(\sqrt{1-b^2 u_o^2}+\sqrt{1-b^2 u_s^2})\Big(-4M+Q-\frac{8Ma^2}{3b^2}\nonumber\\
&-\frac{2a^2Q}{9b^2}-\frac{37M^2Q}{3b^2}\Big)-\frac{1}{2b}\sqrt{1-b^2 u_s^2}\Big(MQu_s-\frac{4a^2Q u_o^2}{9}\nonumber\\
&-\frac{M^2(6u_s+Qu_s^2)}{12}+ (2Q+\frac{4a^2Q}{3b^2})\ln\left[\frac{1}{u_s\, r_0}\right] \Big) \nonumber\\
&-\frac{1}{2b}\sqrt{1-b^2 u_o^2}\Big(MQ u_o-\frac{4a^2Q u_o^2}{9}-\frac{M^2(6u_o+Qu_o^2)}{12}\nonumber\\
&+ (2Q+\frac{4a^2Q}{3b^2})\ln\left[\frac{1}{u_o\, r_0}\right] \Big)+M\left(\frac{21\pi Q}{8b^2}-\frac{3\pi aQ}{b^3}\right)\nonumber\\
& +\frac{M(Qb-aQ)}{b^2}(u_s-u_o)\nonumber\\
&+\frac{MQ(6a-5b)}{2b^3}\left( \sin^{-1}b u_s- \sin^{-1}b u_o\right)\nonumber\\
& +\frac{Q}{b}\left(\ln\left[\frac{b\, u_s}{1-\sqrt{1-b^2 u_s^2}}\right]- \ln\left[\frac{b\, u_o}{1-\sqrt{1-b^2 u_o^2}}\right]\right)\nonumber\\
&+M^2\Big(\frac{15\pi}{4b^2}-\frac{4a\pi}{b^3}-\left(\frac{4}{b}-\frac{4a}{b^2}+\frac{45\pi Q}{8b^2}+\frac{3Q(u_o+u_s)}{2b} \right)\nonumber\\
&\times (u_o-u_s)\Big)+M^2\frac{(16a-15b)}{4b^3} (\sin^{-1}(b u_s)+\sin^{-1}(b u_o))\nonumber\\
& +\mathcal{O}\left(\frac{M^2a^2}{b^4}, \frac{M^2a^2Q}{b^5}\right)	
.\label{Gaussian1}
\end{align}	

Because we considered the source and the observer at the opposite sides to the black hole, we have used $\sin{\phi_o}= b\,u_o$, $\cos\phi_o=-\sqrt{1-b^2u_o^2},\;\; \sin{\phi_s}= b\,u_s, \cos\phi_s=\sqrt{1-b^2u_s^2}$. 

We now calculate the contribution from the geodesic curvature $k_g$, which reads as
\begin{align}
k_g&=-\frac{aQ}{r^3}+\frac{aQ\ln\left[\frac{r}{r_0}\right]}{r^3}+\Big(-\frac{2a}{r^3}-\frac{3aQ}{r^4}+\frac{6aQ\ln\left[\frac{r}{r_0}\right]}{r^4}\Big)M \nonumber\\
&+\mathcal{O}\Big(\frac{M^2a}{r^4}\Big).
\end{align}
It is worth noticing that $k_g$ vanishes for the $a=0$. The contribution from $k_g$ is the path integral along the light curve (from $S$ to $O$), whereas straight lines joining $S$ to $S_{\infty}$ and $O$ to $O_{\infty}$ both have zero geodesic curvature. The line element along the photon orbit is computed from Eqs.~(\ref{lenght}) and (\ref{Orbit}) and reads as
\begin{equation}
d\ell=\frac{1-2Mu+Qu\ln[\frac{1}{u\,r_0}]+a^2u^2}{u^2(1-2Mu+Qu\ln[\frac{1}{u\,r_0}])\big(b+u(a-b)(2M-Q\ln[\frac{1}{u\,r_0}])\big)},
\end{equation}
using the orbits equation from Eq.~(\ref{orbit1}), 
the path integral reads as
\begin{align}
&\int_S^O k_g d\ell =-\frac{3Ma Q}{2b^3}\left( \cos^{-1}b u_s- \cos^{-1}b u_o\right)\nonumber\\
&+\frac{2MaQ}{b^2}(u_o-u_s)-\frac{a}{2b^2}\sqrt{1-b^2\,u_o^2}\Big(4M-\frac{3MQ\,u_o}{2b}\nonumber\\
&-2Q\ln\left[\frac{1}{r_0\,u_o}\right]\Big)-\frac{a}{2b^2}\sqrt{1-b^2\,u_s^2}\Big(4M-\frac{3MQ\,u_s}{2b}\nonumber\\
&-2Q\ln\left[\frac{1}{r_0\,u_s}\right]\Big)+\frac{Qa}{b^2}\ln\Big[\frac{b\,u_o(1-\sqrt{1-b^2\,u_s^2})}{b\,u_s(1+\sqrt{1-b^2\,u_o^2})}\Big]. \label{geodesiccurvature}
\end{align}  
Here, we have adopted the sign convention such that for the prograde (retrograde) photons $d\ell>0$ ($d\ell<0$). The analytical expression for the gravitational deflection angle of light in the weak field limit for the rotating Horndeski black hole can be obtained by using Eqs.~(\ref{Gaussian1}) and (\ref{geodesiccurvature}) in Eq.~(\ref{deflectionangle}). For the infinitely distant source and observer, $u_s\to 0$ and $u_o\to 0$, the deflection angle for prograde photons reads as
\begin{align}
\alpha_D&=\frac{4M}{b}-\frac{Q}{b}-\frac{4Ma}{b^2} +\frac{21\pi M Q}{8b^2} +\frac{15\pi M^2}{4b^2} +\frac{8Ma^2}{3b^3}\nonumber\\
&-\frac{4\pi M^2 a}{b^3}+\frac{2a^2Q}{9b^3}+\frac{37M^2 Q}{3b^3}-\frac{43\pi M a Q}{4b^3}\nonumber\\
&+\frac{Q}{b}\ln\left[\frac{4\, r_0^2}{b^2} \right]+ \frac{aQ}{b^2}\ln\left[\frac{b^2}{4\, r_0^2} \right] -\frac{9\pi M Q}{4b^2}\ln\left[\frac{2b}{ r_0} \right]\nonumber\\
&+ \frac{13\pi M a Q}{2b^3}\ln\left[\frac{b}{ r_0} \right]+ \mathcal{O}\Big(\frac{M^2aQ}{b^4},\frac{Ma^2Q^2}{b^5}\Big), ~\label{deflection}
\end{align}
and for retrograde photons
\begin{align}
\alpha_D&=\frac{4M}{b}-\frac{Q}{b}+\frac{4Ma}{b^2} +\frac{21\pi M Q}{8b^2} +\frac{15\pi M^2}{4b^2} +\frac{8Ma^2}{3b^3}\nonumber\\
&-\frac{4\pi M^2 a}{b^3}+\frac{2a^2Q}{9b^3}+\frac{37M^2 Q}{3b^3}-\frac{19\pi M a Q}{4b^3}\nonumber\\
&+\frac{Q}{b}\ln\left[\frac{4\, r_0^2}{b^2} \right]- \frac{aQ}{b^2}\ln\left[\frac{b^2}{4\, r_0^2} \right] -\frac{9\pi M Q}{4b^2}\ln\left[\frac{2b}{ r_0} \right]\nonumber\\
&- \frac{13\pi M a Q}{2b^3}\ln\left[\frac{b}{ r_0} \right]+ \mathcal{O}\Big(\frac{M^2aQ}{b^4},\frac{Ma^2Q^2}{b^5}\Big),  ~\label{deflection1}
\end{align}
which in the limit $Q=0$ reduce to the Kerr deflection angle $\left.\alpha_D\right|_{\text{Kerr}}$ \cite{Edery:2006hm}. Furthermore, for $a=0$, Eqs.~(\ref{deflection}) and (\ref{deflection1}) define the deflection angle for the nonrotating Horndeski black hole. The black hole rotation breaks the degeneracy in deflection angle, and the prograde and retrograde photons lead to a distinct deflection angle for the same values of parameters, such that the deflection angle for prograde (retrograde)
photons monotonically decrease (increase) with increasing black hole spin. Therefore, the deflection angle around rotating black holes is smaller (larger) than the nonrotating black hole for prograde (retrograde) photons.

Next, we will discuss the astrophysical implication of the rotating Horndeski gravity black hole by calculating the weak gravitational deflection angle. We will model the Sgr A* black hole ($M=4.0\times 10^6 M_{\odot}$, $r_o=d=8.3$ kpc) as the rotating Horndeski gravity black hole and numerically compute the deflection angle and estimate the corrections from the Kerr and nonrotating Horndeski gravity black hole models.
\begin{table}[ht!]
	\centering
	\begin{tabular}{|c|c|c|c|c|c|}
		\hline
		$a/M $  & $Q=$    & $Q=$     & $Q=$     &$ Q= $   &$ Q=$\\
		$$  & $-0.1M$    & $-0.3M$     & $-0.5M$     &$ -0.7M $   &$ -0.9M $\\
		\hline\hline
		0.1& 34.2133& 102.64& 171.067&  239.493& 307.92  \\  
		& (34.214)&  (102.642)&  (171.07)&  (239.498)&  (307.926) \\ \hline
		0.3& 34.2127& 102.638& 171.063& 239.489& 307.914 \\  
		& (34.2146)&  (102.644)&  (171.073)&  (239.502)&  (307.931) \\  \hline
		0.5& 34.212& 102.636& 171.06& 239.484& 307.908 \\  
		& (34.2153)&  (102.646)&  (171.076)&  (239.507)&  (307.937) \\  \hline
		0.7& 34.2114& 102.634& 171.057& 239.48& --- \\  
		& (34.2159)&  (102.648)&  (171.079)&  (239.511)&   \\  \hline
		0.9& 34.2107& ---&  --- & --- &--- \\  
		& (34.2165)&  &  &  &  \\  \hline
		\end{tabular}
		\caption{The corrections in the deflection angle $\delta\alpha_D =\alpha_D-\left.\alpha_D\right|_{\text{Kerr}}$ for Sgr A* with $b=10^4M$, source at $r_s=10^5M$, $r_0=2M$, and varying $Q$ and $a$; $\delta\alpha_D$ is in units of arcsec for prograde (retrograde) photons.}\label{T1}
		\end{table}
		\begin{table}[h!]
		\centering
		\begin{tabular}{|c|c|c|c|c|c|}
		\hline
		$a/M $  & $Q=$    & $Q=$     & $Q=$     &$ Q= $   &$ Q=$\\
		$$  & $-0.1M$    & $-0.3M$     & $-0.5M$     &$ -0.7M $   &$ -0.9M $\\  
		\hline\hline
		0.1& 134.952& 404.855& 674.759& 944.662& 1214.57  \\  
		& (134.971)&  (404.913)&  (674.855)&  (944.797)&  (1214.74) \\ \hline
		0.3& 134.933&404.798& 674.663& 944.528& 1214.39 \\  
		& (134.99)&  (404.97)&  (674.951)&  (944.931)&  (1214.91) \\  \hline
		0.5& 134.913& 404.74& 674.566& 944.393& 1214.22 \\  
		& (135.009)&  (405.028)&  (675.046)&  (945.065)&  (1215.08) \\  \hline
		0.7& 134.894& 404.682& 674.47& 944.258& --- \\  
		& (135.028)&  (405.085)&  (675.142)&  (945.199)&   \\  \hline
		0.9& 134.875& ---&  --- & --- &--- \\  
		& (135.048)&  &  &  &  \\  \hline
	\end{tabular}
	\caption{The corrections in the deflection angle $\delta\alpha_D =\alpha_D-\left.\alpha_D\right|_{\text{Kerr}}$ for Sgr A* with $b=10^3M$, source star S2 at $r_s=1400M$, $r_0=2M$,  and varying $Q$ and $a$; $\delta\alpha_D$ is in units of arcsec for prograde (retrograde) photons.}\label{T2}
\end{table}
\begin{table}[h!]
	\centering
	\begin{tabular}{|c|c|c|c|c|c|}
		\hline	
		$a/M $  & $Q=$    & $Q=$     & $Q=$     &$ Q= $   &$ Q=$\\
		$$  & $-0.1M$    & $-0.3M$     & $-0.5M$     &$ -0.7M $   &$-0.9M $   \\
		\hline\hline
		0.1&	1.14378 & 1.7873 & 2.43082 &3.07435 & 3.71787\\  
		&   (-1.14331)& (-1.78692) & (-2.43053)& (-3.07414)& (-3.71775)\\  \hline
		0.3&   3.43129& 5.36183& 7.29237& 9.22291& 11.1535 \\  
		& (-3.42999)& (-5.36084)& (-7.29169)& (-9.22254)& (-11.1534) \\  \hline
		0.5&   5.71873& 8.93626&  12.1538&  15.3713&  18.5889 \\  
		& (-5.71672)& (-8.93485)& (-12.153)& (-15.3711)& (-18.5892) \\  \hline
		0.7&   8.00612&  12.5106&   17.0151& 21.5196& --- \\ 
		& (-8.00352)& (-12.509)& (-17.0144)& (-21.5198)&  \\  \hline
		0.9&   10.2934 & ---& ---& ---& --- \\  
		& (-10.290)&  & & &  \\  \hline
		\end{tabular}
		\caption{The corrections in the deflection angle $\delta\alpha_D$ from the nonrotating Horndeski gravity black hole $\delta\alpha_D =\left.\alpha_D\right|_{\text{NR}}- \alpha_D$ for Sgr A* with $b=10^4M$ and source at $r_s=10^5M$, $r_0=2M$ ; $\delta\alpha_D$ is in units of mas for prograde (retrograde) photons. }\label{T3}
		\end{table}
		\begin{table}[h!]
		\centering
		\begin{tabular}{|c|c|c|c|c|c|}
		\hline	
		$a/M $  & $Q=$    & $Q=$     & $Q=$     &$ Q= $   &$ Q=$\\
		$$  & $-0.1M$    & $-0.3M$     & $-0.5M$     &$ -0.7M $   &$ -0.9M $   \\
		\hline\hline
		0.1&	0.04307 & 0.062337 & 0.08160 & 0.10086 & 0.12012  \\  
		&   (-0.04280)& (-0.06194) & (-0.081087)& (-0.10023)& (-0.11937)\\  \hline
		0.3&    0.12920& 0.18699& 0.244776& 0.30256& 0.36034 \\  
		& (-0.12842)& (-0.18585)& (-0.24329)& (-0.30072)& (-0.35816) \\  \hline
		0.5&    0.21532& 0.31161& 0.40791& 0.50421& 0.60050 \\  
		& (-3.7766)& (-0.30979)& (-0.40553)& (-0.50126)& (-0.59700) \\  \hline
		0.7&    0.30141& 0.43621& 0.57101& 0.70581& --- \\ 
		& (-0.29971)& (-0.43376)& (-0.56780)& (-0.70185)&  \\  \hline
		0.9&   0.38748 & ---& ---& ---& --- \\  
		& (-0.38539)&  & & &  \\  \hline
	\end{tabular}
	\caption{The corrections in the deflection angle $\delta\alpha_D$ from the nonrotating Horndeski gravity black hole $\delta\alpha_D =\left.\alpha_D\right|_{\text{NR}}- \alpha_D$ for Sgr A* with $b=10^3M$ and source at $r_s=1400M$, $r_0=2M$ ; $\delta\alpha_D$ is in units of arcsec for prograde (retrograde) photons. }\label{T4}
\end{table}

\begin{figure*}
	\begin{center}
		\begin{tabular}{c c}
			\includegraphics[scale=0.7]{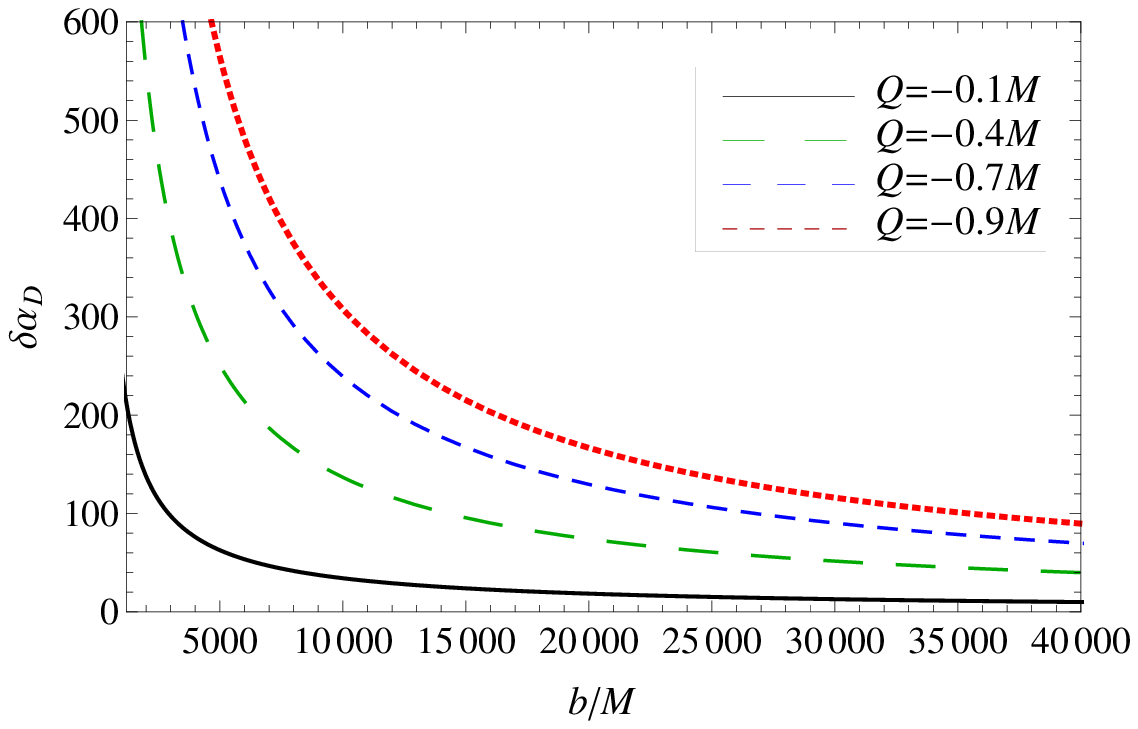}&
			\includegraphics[scale=0.7]{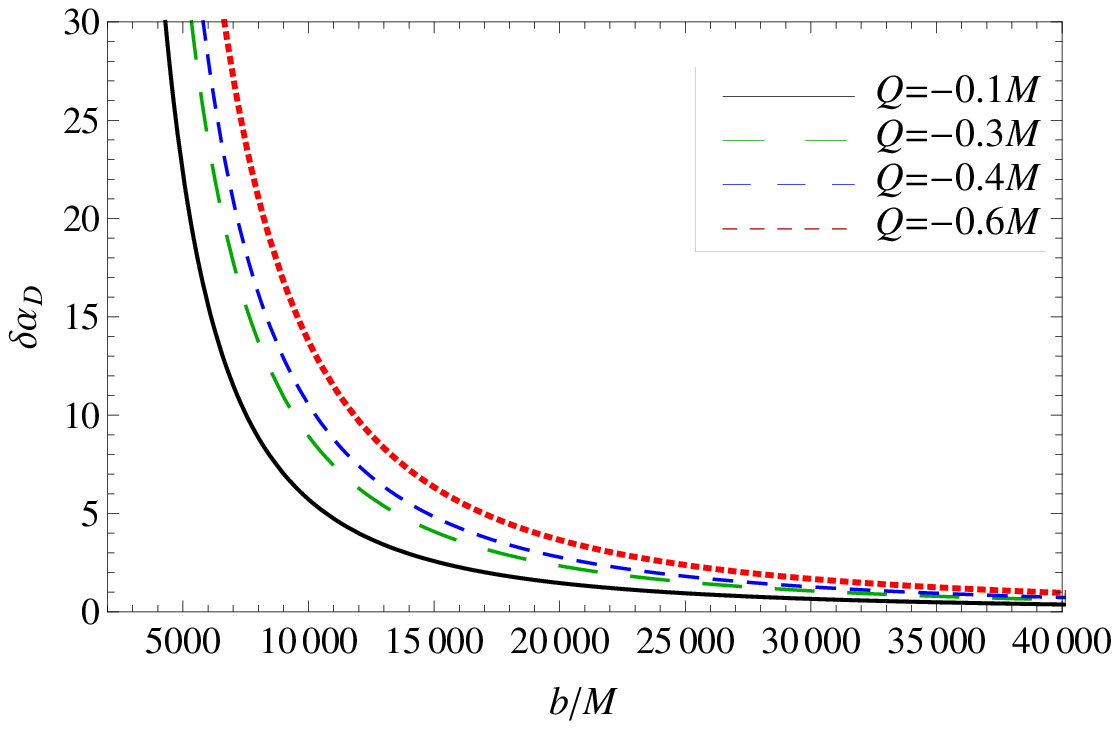}
		\end{tabular}	
	\end{center}
	\caption{\textbf{Left:} Correction in the light deflection angle $\delta\alpha_D$ from Kerr black holes $\delta\alpha_D =\alpha_D-\left.\alpha_D\right|_{\text{Kerr}}$ with $a=0.50M$, and varying $b$; $\delta\alpha_D$ is in units of arcsec. \textbf{Right:} Correction in the light deflection angle $\delta\alpha_D$ from nonrotating Horndeski black holes $\delta\alpha_D =\left.\alpha_D\right|_{\text{NR}}- \alpha_D$ with $a=0.50M$, and varying $b$; $\delta\alpha_D$ is in units of mas.}\label{DefAng}
\end{figure*}

In Tables \ref{T1} \& \ref{T2},  and \ref{T3} \&  \ref{T4} we depicted the calculated corrections in deflection angles $\delta\alpha_D$ for the rotating Horndeski gravity black hole, respectively, from the Kerr and nonrotating Horndeski gravity black holes. Considering the light source at a large distance $r_s=10^5M$ from the black hole and observer at the Earth, the corrections $\delta\alpha_D$ for both prograde and retrograde photons are shown Tables \ref{T1} and \ref{T3}. It is clear from Table \ref{T1} that rotating black holes in the Horndeski gravity cause larger deflection angle than that for the Kerr black hole, and the deflection angle correction $\delta\alpha_D =\alpha_D-\left.\alpha_D\right|_{\text{Kerr}}$ increases with $Q$ for both prograde and retrograde photons, whereas $\delta\alpha_D$ decrease (increase) with $a$ for prograde (retrograde) photons; the $\delta\alpha_D$ are $\mathcal{O}(as)$. This was expected because of the frame-dragging effect; for a fixed value of impact parameter, the prograde photons pass quickly, compared to the retrograde photons, through the black hole gravitational field. Thus the retrograde photons experience larger deflection angle compared to the prograde photons (cf. Table~\ref{T1}). Interestingly, the black hole rotation weakens the gravitational field as the rotating black holes lead to smaller deflection angle compared to the nonrotating black holes for the prograde photons (cf. Table \ref{T3}). On the other hand, the retrograde photons experience larger deflection angle around the rotating black hole in comparison with the nonrotating black holes (cf. Table \ref{T3}). The absolute correction $|\delta\alpha_D| =|(\left.\alpha_D\right|_{\text{NR}}- \alpha_D)|$ from the nonrotating Horndeski gravity black holes monotonically increases with $a$ and  $Q$ for the both prograde and retrograde photon; and are of $\mathcal{O}(mas)$ for $b=10^4M$(cf. Table \ref{T3}). 

Next, we consider the S2 star as the source star that lives in the bulge of the Milky Way Galaxy, and in May 2018 that approached the closest distance to Sgr A*, $r_s=1400\,M$, therefore, the finite-distance corrections to the deflection angle cannot be neglected. We calculate the light deflection angle and estimate the corrections $\delta\alpha_D$, which are summarized in Tables \ref{T2} and \ref{T4}. The order of correction is arcsec, which is well within the resolution of today's observational facilities. It can be inferred from Table \ref{T2} that both the charge $Q$ and spin $a$ increase the correction in the light deflection angle for retrograde photons. Whereas, $\delta\alpha_D$ decreases with $a$ for prograde photons. The qualitative behavior is exactly same as in Table~\ref{T1}. It is worth mentioning here that for the fixed value of impact parameter, the $\delta\alpha_D$ is smaller for a finite distant source $r_s=1400\,M$ in comparison to a infinite distant source $r_s\to\infty$. In Fig.~\ref{DefAng}, we have depicted how the correction in the deflection angle $\delta\alpha_D$ from the Kerr black hole and nonrotating Horndeski gravity black hole vary with dimensionless impact parameter $b/M$ for different values of $Q$. With the increasing impact parameter $b$, the $\delta\alpha_D$ decreases.

\begin{figure}
	\includegraphics[scale=0.75]{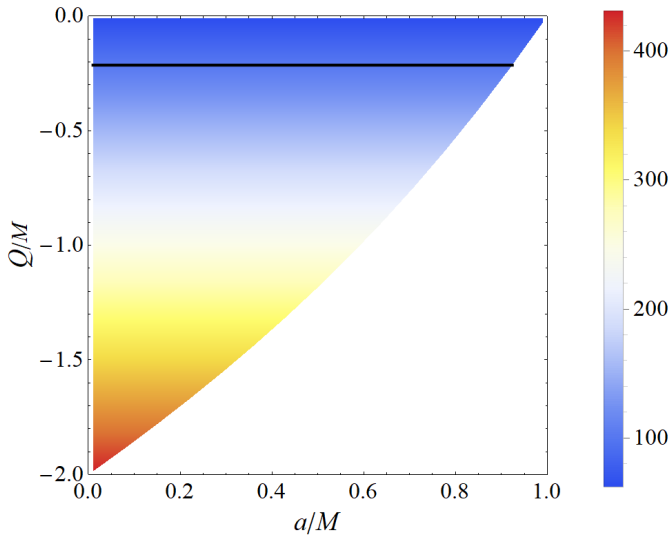}
	\caption{Angular separation $(\mu\,as)$ between two images for Sgr A* black hole with source star at $r_s=1400M$. The black solid line corresponds to $100\,\mu as$.}\label{Img1}
\end{figure}

In fact, Sgr A*, acting as a convex gravitational lens, is able to deflect the light rays emitted by the S-stars from their trajectories, affecting their measured image positions. Angular radius of Sgr A* black hole as seen from the earth is \cite{Gillessen:2009ht} \begin{align}
\theta_s&=\frac{2GM}{d c^2},\\
&=10.1954\,\mu as.
\end{align}
For the rotating Horndeski black hole, the angular separation between two lensed images (formed by prograde and retrograde photons) is $\mathcal{O}(\mu\,as)$ and increases with $Q$ Fig.~\ref{Img1}. The modern interfermometry instruments such as, PRIMA, ASTRA and EHT are conceived to achieve an astrometric accuracy of $\mathcal{O}(\mu as)$ and thereby are capable to observe this image displacement. Currently, GRAVITY, has the angular resolution of about 3 $mas$ at K band $(1.95–2.45)\mu m$. 
Considering the source star at $r_s=1400M$, for the Kerr black hole the image angular separation is $60.10\mu$as and is almost immune to the spin parameter, however, for Horndeski black holes it strongly depends on $Q$. It is worthwhile to notice that the calculated shadow angular diameter for the Sgr A* black hole is $\sim 50\,\mu $as. In summary, black holes in Horndeski gravity lead to larger angular separation between the lensed images compared to that in the general relativity. Although, it is in principle possible to resolve the two images for the Sgr A* black hole by the EHT, it is outside the reach of the current best ability of GRAVITY. Constraints on the Horndeski gravity black holes are derived from the M87* black hole shadow results from the EHT \cite{Afrin:2021wlj}.
	
\section{Conclusions}\label{sec5}
The modified theories of gravity enrich the dynamical field content of general relativity by including scalar fields in the latter, which constitute additional degrees of freedom. One of the most renowned scalar-tensor theories is Horndeski gravity. It is the most general four-dimensional scalar-tensor theory with equations of motion containing second-order derivatives of the dynamical fields. The Horndeski action involves four arbitrary functions of the canonical kinetic term $\chi$, denoted by $Q_{i},\; i = 2,\cdot \cdot \cdot, 5$. Recently, it was shown that when $Q_5=0$, the resulting quartic Horndeski theory of gravity admits spherically symmetric hairy black holes \cite{Bergliaffa:2021diw}. We derived the rotating counterpart of this solution, i.e., a rotating hairy black hole using revised NJA. The derived Kerr-like black hole has an additional Horndeski charge parameter  $Q$ besides mass $M$ and spin parameter $a$. The scalar field produces a hair that changes the structure of the rotating black hole through an additional term $Q$ in the metric (\ref{rotbhtr}), which is asymptotically flat. The rotating hairy black hole metric (\ref{rotbhtr}) can represent black holes with Cauchy and event horizons, an extreme black hole or naked singularity, depending on the choice of the parameters. Constraints on the value of $Q$ are derived for which rotating Horndeski gravity black holes with a given value of spin $a$ possess two distinct horizons. Despite the complicated rotating hairy black hole metric (\ref{rotbhtr}), using the Komar prescription, we found exact expressions for conserved mass $M_{\text{eff}}$ and angular momentum $J_{\text{eff}}$, valid at any radial distance. Furthermore, the hair parameter $Q$ significantly altered these conserved quantities compared with those for the Kerr black hole discovered in the limit $Q=0$. The null Killing vector $\chi^{\mu}$ at the event horizon leads to the corresponding Komar conserved quantity $\mathcal{K}_{\chi}$, which is twice the product of entropy and temperature of the black hole $\mathcal{K}_{\chi}=2S_+T_+$ and hence satisfies the Smarr formula.

We have also derived the light deflection angle in the weak field limit using the Gauss-Bonnet theorem and considering the source and observer at finite distances from the black hole. The corrections in the deflection angle from the Kerr black holes and the non-rotating Horndeski gravity black holes are calculated as an explicit function of the distances of source and observer. For rotating black holes, the deflection angle also depends on the photon angular momentum relative to the black hole, such as the prograde (retrograde) photons lead to a smaller (larger) deflection angle than that for the non-rotating black hole. We have modelled the Sgr A* black hole with the rotating Horndeski gravity black hole and shown that the deflection angle increases with $|Q|$ for prograde and retrograde photons. For the S2 source star at $r_s=1400 M$, the correction in deflection angle $|\delta\alpha_D| =|(\left.\alpha_D\right|_{\text{NR}}- \alpha_D)|$ from the nonrotating Horndeski gravity black holes for light impact parameter $b=1300M$ are up to 75as, which can be measured with the current observational facilities.  To conclude, a rotating hairy solution in Horndeski theories may give new opportunities to test these theories against astrophysical observations. It would also be interesting to consider the M87* black hole's shadow observational results by EHT to put constraints on the parameter $Q$. The problem of the formation of critical caustic curves and the consequent appearance of multiple images in the strong-field deflection limit will be addressed elsewhere.
Further, the AdS background for these rotating hairy black holes should provide exciting phase structure and critical phenomena. More severe constraints are likely to be expected by considering the surrounding accretion disk. The stability of the obtained rotating solution against the scalar perturbations in the gravitational wave observational data may be helpful. Such investigations have a clear astrophysical relevance; we hope to report on these issues in the future. 

\begin{acknowledgements} 
	R.K.W. and S.D.M. thank the NRF and the University of KwaZulu-Natal for continued support. S.D.M acknowledges that this work is based upon research supported by the South African Research Chair Initiative of the Department of Science and Technology and the National Research Foundation. S.G.G. would like to thank the Science and Engineering Research Board, Department of Science and Technology, India for the project No. CRG/2021/005771, and Shaqat Ul Islam for help in plots.
\end{acknowledgements}


\begin{thebibliography}{99}
	
	
	\bibitem{Schwarzschild:1916uq}
	K.~Schwarzschild,
	Sitzungsber. Preuss. Akad. Wiss. Berlin (Math. Phys. ) \textbf{1916}, 189 (1916).
	
	\bibitem{Kerr:1963ud} 
	R.~P.~Kerr,
	Phys.\ Rev.\ Lett.\  {\bf 11}, 237 (1963).
	
	\bibitem{Newman:1965tw} 
	E. T. Newman and A. I. Janis, J. Math. Phys. {\bf 6}, 915
	(1965).
	
	
	\bibitem{Akiyama:2019cqa}
	K.~Akiyama \textit{et al.} [Event Horizon Telescope],
	Astrophys. J. Lett. \textbf{875}, L1 (2019). 
	
	\bibitem{Akiyama:2019bqs}
	K.~Akiyama \textit{et al.} [Event Horizon Telescope],
	Astrophys. J. Lett. \textbf{875}, L4 (2019).
	
	
	\bibitem{EventHorizonTelescope:2020qrl}
	D.~Psaltis \textit{et al.} [Event Horizon Telescope],
	Phys. Rev. Lett. \textbf{125}, no.14, 141104 (2020).
	
	\bibitem{Bambi1} 
	C.~Bambi, K.~Freese, S.~Vagnozzi and L.~Visinelli,
	Phys.\ Rev.\ D {\bf 100}, 044057 (2019);
	S.~Vagnozzi and L.~Visinelli,
	Phys.\ Rev.\ D {\bf 100}, 024020 (2019);
	R.~Kumar, S.~G.~Ghosh and A.~Wang,
	Phys.\ Rev.\ D {\bf 100}, 124024 (2019);
	P.~V.~P.~Cunha, C.~A.~R.~Herdeiro and E.~Radu,
	Universe {\bf 5}, 220 (2019);
	R.~Kumar, B.~P.~Singh and S.~G.~Ghosh,
	Annals Phys. \textbf{420}, 168252 (2020);
	I.~Banerjee, S.~Chakraborty and S.~SenGupta,
	Phys. Rev. D \textbf{101}, 041301 (2020);
	I.~Banerjee, S.~Sau and S.~SenGupta,
	Phys. Rev. D \textbf{101}, 104057 (2020);
	A.~Allahyari, M.~Khodadi, S.~Vagnozzi and D.~F.~Mota,
	JCAP \textbf{02}, 003 (2020);
	J.~W.~Moffat and V.~T.~Toth,
	Phys.\ Rev.\ D {\bf 101}, 024014 (2020);
	R.~Kumar, A.~Kumar and S.~G.~Ghosh,
	Astrophys. J. \textbf{896}, 89 (2020);
	R.~Kumar and S.~G.~Ghosh,
	Astrophys. J. \textbf{892}, 78 (2020);
	M.~Wielgus \textit{et al.},
	Astrophys. J. \textbf{901}, 67 (2020);
	M.~Khodadi, G.~Lambiase and D.~F.~Mota,
	JCAP \textbf{09}, 028 (2021);
	M.~Khodadi, A.~Allahyari, S.~Vagnozzi and D.~F.~Mota,
	JCAP \textbf{09}, 026 (2020);
	M.~Afrin, R.~Kumar and S.~G.~Ghosh,
	Mon. Not. Roy. Astron. Soc. \textbf{504}, 5927 (2021);
	S.~G.~Ghosh, R.~Kumar and S.~U.~Islam,
	JCAP \textbf{03}, 056 (2021);
	R.~Kumar and S.~G.~Ghosh,
	Class. Quant. Grav. \textbf{38}, 8 (2021).
	
	\bibitem{Medeiros:2019cde}
	L.~Medeiros, D.~Psaltis and F.~\"Ozel,
	Astrophys. J. \textbf{896}, 7 (2020).
	
	\bibitem{EventHorizonTelescope:2021dqv}
	P.~Kocherlakota \textit{et al.} [Event Horizon Telescope],
	Phys. Rev. D \textbf{103}, 104047 (2021)
	
	\bibitem{weinberg} S. Weinberg, Rev. Mod. Phys. \textbf{61} 1, 1989.
	
	\bibitem{Lee:2022cyh}
	B.~H.~Lee, W.~Lee, E.~\'O.~Colg\'ain, M.~M.~Sheikh-Jabbari and S.~Thakur,
	JCAP \textbf{04}, no.04, 004 (2022)
	
	\bibitem{Heisenberg:2022gqk}
	L.~Heisenberg, H.~Villarrubia-Rojo and J.~Zosso,
	[arXiv:2202.01202 [astro-ph.CO]].
	
	\bibitem{Heisenberg:2022lob}
	L.~Heisenberg, H.~Villarrubia-Rojo and J.~Zosso,
	[arXiv:2201.11623 [astro-ph.CO]].
	
	
	\bibitem{Damour}T. Damour and G. Esposito-Farese, Class. Quant. Grav. \textbf{9}, 2093 (1992)
	
	\bibitem{Horndeski:1974wa}
	G.~W.~Horndeski,
	Int. J. Theor. Phys. \textbf{10}, 363 (1974).
	
	
	\bibitem{kase} 
	R. Kase, S. Tsujikawa, Int. J. Mod. Phys. D \textbf{28}, 1942005 (2019).
	
	\bibitem{koba} 
	T. Kobayashi, Rept. Prog. Phys. \textbf{82}, 086901  (2019). 
	
	\bibitem{Ostrogradsky} 
	M. Ostrogradsky, Mem. Ac. St. Petersbourg VI 4  385 (1850).
	
	\bibitem{Lu:2020iav}
	H.~Lu and Y.~Pang,
	Phys. Lett. B \textbf{809}, 135717 (2020).
	
	\bibitem{VanAcoleyen:2011mj}
	K.~Van Acoleyen and J.~Van Doorsselaere,
	Phys. Rev. D \textbf{83}, 084025 (2011).
	
	\bibitem{Charmousis:2014mia}
	C.~Charmousis,
	Lect. Notes Phys. \textbf{892}, 25-56 (2015).
	
	\bibitem{Rinaldi:2012vy}
	M.~Rinaldi,
	Phys. Rev. D \textbf{86}, 084048 (2012).
	
	\bibitem{Babichev:2014fka}
	E.~Babichev and A.~Fabbri,
	JHEP \textbf{07}, 016 (2014).
	
	\bibitem{Anabalon:2013oea}
	A.~Anabalon, A.~Cisterna and J.~Oliva,
	Phys. Rev. D \textbf{89}, 084050 (2014).
	
	\bibitem{Cisterna:2014nua}
	A.~Cisterna and C.~Erices,
	Phys. Rev. D \textbf{89}, 084038 (2014).
	
	\bibitem{Bravo-Gaete:2014haa}
	M.~Bravo-Gaete and M.~Hassaine,
	Phys. Rev. D \textbf{90}, 024008 (2014).
	
	\bibitem{Herdeiro:2015waa}
	C.~A.~R.~Herdeiro and E.~Radu,
	Int. J. Mod. Phys. D \textbf{24}, 1542014 (2015).
	
	\bibitem{Herdeiro:2014goa}
	C.~A.~R.~Herdeiro and E.~Radu,
	Phys. Rev. Lett. \textbf{112}, 221101 (2014).
	
	\bibitem{Gao:2021luq}
	Y.~X.~Gao and Y.~Xie,
	Phys. Rev. D \textbf{103}, no.4, 043008 (2021).
	\bibitem{Herdeiro:2016tmi}
	C.~Herdeiro, E.~Radu and H.~R\'unarsson,
	Class. Quant. Grav. \textbf{33}, 154001 (2016).
	
	\bibitem{Hui:2012qt}
	L.~Hui and A.~Nicolis,
	Phys. Rev. Lett. \textbf{110}, 241104 (2013).
	
	
	\bibitem{Babichev:2017guv}
	E.~Babichev, C.~Charmousis and A.~Leh\'ebel,
	JCAP \textbf{04}, 027 (2017).
	
	\bibitem{Nicolis:2008in}
	A.~Nicolis, R.~Rattazzi and E.~Trincherini,
	Phys.\ Rev.\ D {\bf 79},  064036 (2009).
	
	\bibitem{Kase:2018aps}
	R.~Kase and S.~Tsujikawa,
	Int. J. Mod. Phys. D \textbf{28}, 1942005 (2019).
	
	\bibitem{Amendola:2018ltt}
	L.~Amendola, D.~Bettoni, G.~Dom\`enech and A.~R.~Gomes,
	JCAP \textbf{06}, 029 (2018).

	\bibitem{Bergliaffa:2021diw}
	S.~E.~P.~Bergliaffa, R.~Maier and N.~d.~Silvano,
	arXiv:2107.07839 [gr-qc].
	
		
	\bibitem{Kumar:2021cyl}
	J.~Kumar, S.~U.~Islam and S.~G.~Ghosh,
	Eur. Phys. J. C \textbf{82}, 443 (2022).
	
	\bibitem{Virbhadra:1998dy}
	K.~S.~Virbhadra, D.~Narasimha and S.~M.~Chitre,
	Astron. Astrophys. \textbf{337}, 1-8 (1998).
	
	
	\bibitem{Newman:1965my}
	E.~T.~Newman, R.~Couch, K.~Chinnapared, A.~Exton, A.~Prakash and R.~Torrence,
	J. Math. Phys. \textbf{6}, 918-919 (1965).
	
	
	\bibitem{Azreg-Ainou:2014pra} 
	M.~Azreg-A\"\i{}nou,
	Phys. Rev. D \textbf{90}, 064041 (2014).
	
	\bibitem{Azreg-Ainou:2014aqa} 
	M.~Azreg-Aïnou, 
	Eur. Phys. J. C {\bf 74},  2865 (2014).
	
	\bibitem{Johannsen:2011dh}
	T.~Johannsen and D.~Psaltis,
	Phys. Rev. D \textbf{83}, 124015 (2011).
	
	\bibitem{Bambi:2013ufa}
	C.~Bambi and L.~Modesto,
	Phys.\ Lett.\ B \textbf{721}, 329 (2013).
	
	\bibitem{Ghosh:2014pba}
	S.~G.~Ghosh,
	Eur. Phys. J. C \textbf{75}, 532 (2015).
	
	\bibitem{Moffat:2014aja} 
	J.~W.~Moffat,
	Eur.\ Phys.\ J.\ C {\bf 75}, 175 (2015).
	
	\bibitem{Kumar:2020owy}
	R.~Kumar and S.~G.~Ghosh,
	JCAP \textbf{07},  053 (2020).
	
	\bibitem{Kumar:2020hgm}
	R.~Kumar, S.~G.~Ghosh and A.~Wang,
	Phys. Rev. D \textbf{101}, 104001 (2020).
	
	\bibitem{Kumar:2017qws}
	R.~Kumar and S.~G.~Ghosh,
	Eur. Phys. J. C \textbf{78}, 750 (2018).
	
	\bibitem{Brahma:2020eos}
	S.~Brahma, C.~Y.~Chen and D.~h.~Yeom,
	Phys. Rev. Lett. \textbf{126}, 181301 (2021).
	
	\bibitem{Hawking:1971vc} 
	S.~W.~Hawking,
	Commun.\ Math.\ Phys.\  {\bf 25}, 152 (1972).
	
	\bibitem{he} 
	S.W.~Hawking and G.F.R.~Ellis, {\it The
		Large Scale Structure of Spacetime} Cambridge University Press,
	Cambridge (1973).
	
	\bibitem{Poisson:2009pwt} 
	E.~Poisson,
	\textit{A Relativist's Toolkit: The Mathematics of Black-Hole Mechanics} Cambridge University Press, Cambridge, England, (2004).
	
	
	\bibitem{Neves:2014aba}
	J.~C.~S.~Neves and A.~Saa,
	Phys. Lett. B \textbf{734}, 44 (2014).
	
	
	
	\bibitem{Brown:2018hym}
	P.~J.~Brown, C.~J.~Fewster and E.~A.~Kontou,
	Gen. Rel. Grav. \textbf{50}, 121 (2018).
	
	\bibitem{Klinkhammer:1991ki}
	G.~Klinkhammer,
	Phys. Rev. D \textbf{43}, 2542 (1991).
	
	\bibitem{Ford:2000xg}
	L.~H.~Ford and T.~A.~Roman,
	Phys. Rev. D \textbf{64}, 024023 (2001).
	
	
	\bibitem {Chandrasekhar:1992}
	S.~ Chandrasekhar, 
	{\it The Mathematical Theory of Black Holes}, Oxford University Press, New York, (1992).
	
	\bibitem{Wald} 
	R. M. Wald, \textit{General Relativity}, University of Chicago Press, Chicago, (1984).
	
	\bibitem{Komar:1958wp} 
	A.~Komar,
	Phys.\ Rev.\  {\bf 113}, 934 (1959).
	
	\bibitem{Chen:2008ra} 
	S.~Chen, B.~Wang and R.~Su,
	Phys.\ Rev.\ D {\bf 77}, 124011 (2008).
	
	\bibitem{Sekiwa:2006qj} 
	Y.~Sekiwa,
	Phys.\ Rev.\ D {\bf 73}, 084009 (2006).
	
	\bibitem{Smarr:1972kt} 
	L.~Smarr,
	Phys.\ Rev.\ Lett.\  {\bf 30}, 71 (1973).
	
	\bibitem{Bardeen:1973gs}
	J.~M.~Bardeen, B.~Carter and S.~W.~Hawking,
	Commun. Math. Phys. \textbf{31}, 161 (1973)
	
	\bibitem{Sahabandu:2005ma} 
	C.~Sahabandu, P.~Suranyi, C.~Vaz and L.~C.~R.~Wijewardhana,
	Phys.\ Rev.\ D {\bf 73}, 044009 (2006).
	
	\bibitem{Cai:2003kt}
	R.~G.~Cai,
	Phys.\ Lett.\ B {\bf 582}, 237 (2004).
	\bibitem{Altamirano:2014tva}
	N.~Altamirano, D.~Kubiznak, R.~B.~Mann and Z.~Sherkatghanad,
	Galaxies \textbf{2}, 89 (2014).
	\bibitem{Carlip:2003ne}
	S.~Carlip and S.~Vaidya,
	Class. Quant. Grav. \textbf{20}, 3827 (2003). 
	\bibitem{Ghosh:2014pga} 
	S.~G.~Ghosh, U.~Papnoi and S.~D.~Maharaj,
	Phys.\ Rev.\ D {\bf 90}, 044068 (2014);
	S.~G.~Ghosh, D.~V.~Singh and S.~D.~Maharaj,
	Phys.\ Rev.\ D {\bf 97}, 104050 (2018);
	S.~G.~Ghosh,
	Class.\ Quant.\ Grav.\  {\bf 35}, 085008 (2018). 
	
	\bibitem{Virbhadra:2008ws}
	K.~S.~Virbhadra,
	Phys. Rev. D \textbf{79}, 083004 (2009).
	
	\bibitem{Virbhadra:2007kw}
	K.~S.~Virbhadra and C.~R.~Keeton,
	Phys. Rev. D \textbf{77}, 124014 (2008).
	
	\bibitem{Virbhadra:2002ju}
	K.~S.~Virbhadra and G.~F.~R.~Ellis,
	Phys. Rev. D \textbf{65}, 103004 (2002).
	
	\bibitem{Gibbons:2008rj} 
	G.~W.~Gibbons and M.~C.~Werner,
	Class. Quant. Grav. {\bf 25}, 235009 (2008).
	
	\bibitem{carmo} M. P. Do Carmo, \textit{Differential Geometry of Curves and Surfaces} (Prentice-Hall, New Jersey, 1976).
	
	\bibitem{Ishihara:2016vdc} 
	A.~Ishihara, Y.~Suzuki, T.~Ono, T.~Kitamura, and H.~Asada,
	Phys.\ Rev.\ D {\bf 94},  084015 (2016); 
	A.~Ishihara, Y.~Suzuki, T.~Ono and H.~Asada,
	Phys.\ Rev.\ D {\bf 95}, 044017 (2017).
	
	\bibitem{Ono:2017pie} 
	T.~Ono, A.~Ishihara and H.~Asada,
	Phys.\ Rev.\ D {\bf 96}, 104037 (2017).
	
	\bibitem{Crisnejo1:2018uyn} 
	G.~Crisnejo and E.~Gallo,
	Phys.\ Rev.\ D {\bf 97}, 124016 (2018);
	A.~\"{O}vg\"{u}n,
	Phys.\ Rev.\ D {\bf 98}, 044033 (2018);
	A.~\"{O}vg\"{u}n, I.~Sakalli and J.~Saavedra,
	J. Cosmol. Astropart. Phys. {\bf 1810}, 041 (2018);
	A.~\"{O}vg\"{u}n,
	Phys.\ Rev.\ D {\bf 99}, 104075 (2019);
	W.~Javed, j.~Abbas and A.~\"{O}vg\"{u}n,
	Phys.\ Rev.\ D {\bf 100}, 044052 (2019);
	W.~Javed, R.~Babar and A.~\"{O}vg\"{u}n,
	Phys.\ Rev.\ D {\bf 100}, 104032 (2019);
	W.~Javed, J.~Abbas and A.~\"{O}vg\"{u}n,
	Eur.\ Phys.\ J.\ C {\bf 79}, 694 (2019);
	G.~Crisnejo, E.~Gallo and J.~R.~Villanueva,
	Phys.\ Rev.\ D {\bf 100}, 044006 (2019);
	G.~Crisnejo, E.~Gallo and A.~Rogers,
	Phys.\ Rev.\ D {\bf 99}, 124001 (2019);
	T.~Zhu, Q.~Wu, M.~Jamil and K.~Jusufi,
	Phys.\ Rev.\ D {\bf 100}, 044055 (2019).
	
	\bibitem{Ishihara:2016sfv}
	A.~Ishihara, Y.~Suzuki, T.~Ono and H.~Asada,
	Phys. Rev. D \textbf{95}, 044017 (2017).
	
	\bibitem{Werner:2012rc} 
	M.~C.~Werner,
	Gen.\ Relativ.\ Gravit.\  {\bf 44}, 3047 (2012).
	
	\bibitem{Crisnejo:2019ril}
	G.~Crisnejo, E.~Gallo and K.~Jusufi,
	Phys.\ Rev.\ D {\bf 100}, 104045 (2019).	
	
	\bibitem{Edery:2006hm}
	A.~Edery and J.~Godin,
	Gen. Rel. Grav. \textbf{38}, 1715 (2006).	
	
	\bibitem{Gillessen:2009ht}
	S.~Gillessen, F.~Eisenhauer, T.~K.~Fritz, H.~Bartko, K.~Dodds-Eden, O.~Pfuhl, T.~Ott and R.~Genzel,
	Astrophys. J. Lett. \textbf{707}, L114-L117 (2009).
	
	\bibitem{Afrin:2021wlj}
	M.~Afrin and S.~G.~Ghosh,
	Astrophys. J. \textbf{932}, 51 (2022).
	
\end{thebibliography}
\end{document}